\documentstyle{mn2e}

\include{psfig}

\title{The DA+dMe eclipsing binary EC13471--1258: its cup runneth over ... 
       just \thanks{}}

\author[D. O'Donoghue {\it et al.}]
{D. O'Donoghue$^{1,2}$, C. Koen$^1$, D. Kilkenny$^1$, R.S. Stobie$^1$, 
D. Koester$^3$, M.S. Bessell$^4$, \and N. Hambly$^5$, H. MacGillivray$^5$\\
$^{1}$South African Astronomical Observatory, PO Box 9, Observatory 7935, 
      Cape, South Africa \\
$^{2}$Dept. of Astronomy, University of Cape Town, Rondebosch 7700, South 
      Africa\\
$^{3}$Institut f\"ur Theoretische Physik und Astrophysik, Universit\"at
      Kiel, 24098 Kiel, Germany\\
$^{4}$Research School of Astronomy and Astrophysics, Institute of Advanced
      Studies, Australian National University, Cotter Road, \\ Weston Creek,
      Canberra ACT 2611, Australia\\
$^{5}$Wide Field Astronomy Unit, Institute for Astronomy, University of
      Edinburgh, Blackford Hill, Edinburgh, EH9 3HJ, UK}

\date{Accepted \ \ \ \ \ \ \ \ \ \ \ \ \ \ \ \ \ \ \ \ \ \ \ \ \ \ \ \ \ \ \ Received}
\begin{document}

\maketitle

\begin{abstract}
The optical spectrum and light curve of EC13471--1258 shows that it is an
eclipsing binary with an orbital period of 3$^h$ 37$^m$ comprising a DA
white dwarf and a dMe dwarf. Total eclipses of the white dwarf are observed 
lasting 14 min, with the partial phases lasting 54 s. On one occasion, two 
pre-eclipse dips were seen. Timings of the eclipses over ten years show 
jitter of up to 12 s. Flares from the M dwarf are regularly observed. The 
M dwarf also shows a large amplitude ellipsoidal modulation in the V band 
light curve. The component stars emit almost equal amounts of light 
at 5500 \AA.

HST STIS spectra show strong Lyman alpha absorption with weak metal lines
of C I,II and Si II superimposed. Model atmosphere analysis yielded an
effective temperature of $14220 \pm 300$ K and log g of $8.34 \pm 0.20$ for 
the white dwarf with these errors strongly correlated. Its metal abundance 
is 1/30th solar with an uncertainty of 0.5 dex, and it is rapidly rotating 
with ${\rm V_1\ sin\ i = 400 \pm 100}$ km s$^{-1}$. The white dwarf also 
shows radial velocity variations with a semi-amplitude of $138 \pm 10$ km 
s$^{-1}$. The gravitational redshift of the white dwarf was measured: 62 
km s$^{-1}$.

From optical spectroscopy the spectral type of the M dwarf was found to be 
M3.5-M4, its temperature $3100 \pm 75$ K, its rotational velocity $140 \pm 
10$ km s$^{-1}$, its radial velocity semi-amplitude $266 \pm 5$ km s$^{-1}$, 
its mean $V-I$ colour 2.86 and its absolute V magnitude 11.82. Intriguingly, 
its metal abundance is normal solar.

The H$\alpha$ emission line shows at least two distinct components, one of
which is uniformly distributed around the centre of mass of the M dwarf and
provided the estimate of the rotational velocity of the M dwarf. The other 
arises from the other side of the binary centre of mass, well within the 
white dwarf Roche lobe. This behaviour is confirmed by Doppler tomography 
which shows the presence of two distinct velocity components within the
primary Roche lobe. The interpretation of these features is uncertain. 
Variations in strength of the components with binary phase can be attributed 
to optical thickness in the Balmer lines. Similar behaviour is seen in the
observations of the other Balmer emission lines, although with poorer 
signal-to-noise. Flares in H$\alpha$ were observed and are consistent with 
arising from the vicinity of the M dwarf. 

Dynamical solutions for the binary are discussed and yield an inclination of 
$75.5\pm 2.0^{\rm o}$, a white dwarf mass and radius of $0.78\pm0.04$ M$_\odot$ 
and $0.011\pm0.01$ R$_\odot$, and an M dwarf mass and radius of $0.43\pm0.04$ 
M$_\odot$ and $0.42\pm0.02$ R$_\odot$. These parameters are consistent with
the Wood (1995) mass-radius relation for white dwarfs and the Clemens et
al. (1998) mass-radius relation for M dwarfs; we argue that the M dwarf {\em 
just} fills its Roche lobe. The radius of the white dwarf and the model fit
imply a distance of $48\pm5$ pc and an absolute V magnitude of 11.74.

The rapid rotation of the white dwarf strongly suggests that the system has
undergone mass transfer in the past, and implies that it is a hibernating
cataclysmic variable. The M dwarf shows the properties expected of
secondaries in cataclysmic variables: chromospheric activity and angular
momentum loss.

\end{abstract}


\section{Introduction and system overview}

\renewcommand{\thefootnote}{\fnsymbol{footnote}}

\footnotetext[1]{
Based on observations made with the NASA/ESA Hubble Space Telescope, 
obtained at the Space Telescope Science Institute, which is operated 
by the Association of Universities for Research in Astronomy, Inc., 
under NASA contract NAS 5-26555. These observations are associated 
with program 7744.}

Stellar evolution is one of the most secure theories in modern
astrophysics; masses and radii of stars are fundamental parameters
of the stars to which this theory applies. When stars are found in
interacting binaries, additional complexities are introduced; results 
from single star evolution are often applied for want of any alternative 
approach. Systems which offer the prospect of testing these prejudices 
are therefore worthy of some scrutiny. 

Prominent among the interacting binaries are cataclysmic variables: 
short period binaries comprising a Roche lobe filling late K or M dwarf 
and a white dwarf. If the white dwarf's magnetic field is small, the mass 
that is transferred from the cool star to the white dwarf forms an accretion 
disc. The energy generated by the accretion almost always far exceeds the 
luminosity of the component stars, affording a valuable laboratory for 
studying accretion processes. However, masses, radii, temperatures and other 
fundamental properties of the component stars are usually very difficult to 
determine because of the glare of the accretion light. 

The progenitors of cataclysmic variables are detached, short period 
DA+dM binaries which have emerged from common envelope evolution. This
process is thought to give rise to all very short period binaries and
although there has been a lot of theoretical work in this area, 
comprehensive observational tests of the theory are required.

In this paper we report an observational study of a DA + dMe eclipsing 
binary with an orbital period of 3$^{\rm h}37^{\rm m}$. As we will show, 
this system is almost a cataclysmic variable: there is a little mass 
transfer taking place, but the component stars can nevertheless be studied 
with relative ease, their fundamental properties derived and compared with 
theory.

   \begin{figure} 
   \begin{center}
   \begin{tabular}{l}
   \psfig{figure=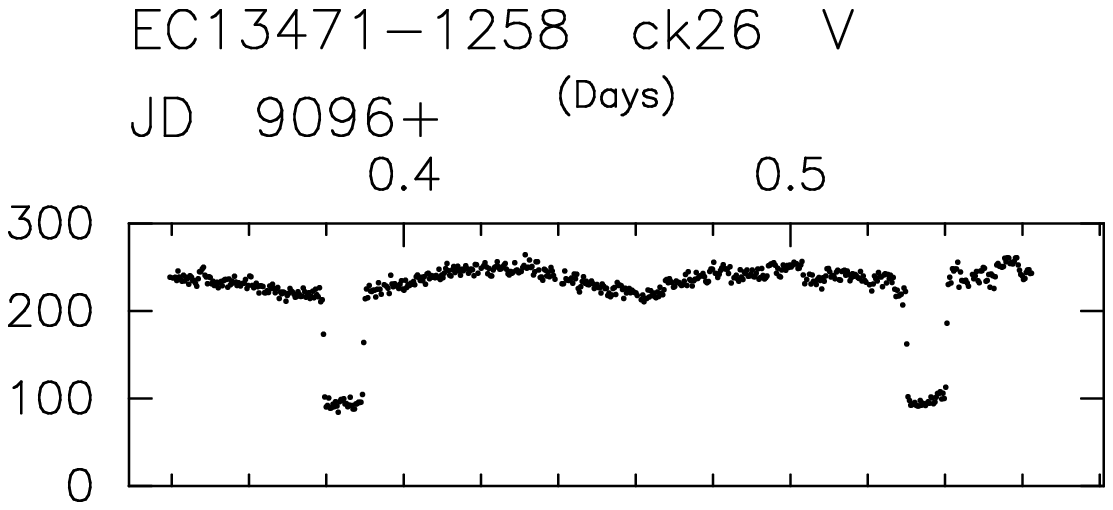,height=3.5cm,angle=0} \\
   \\
   \end{tabular}
   \begin{tabular}{c}
   \psfig{figure=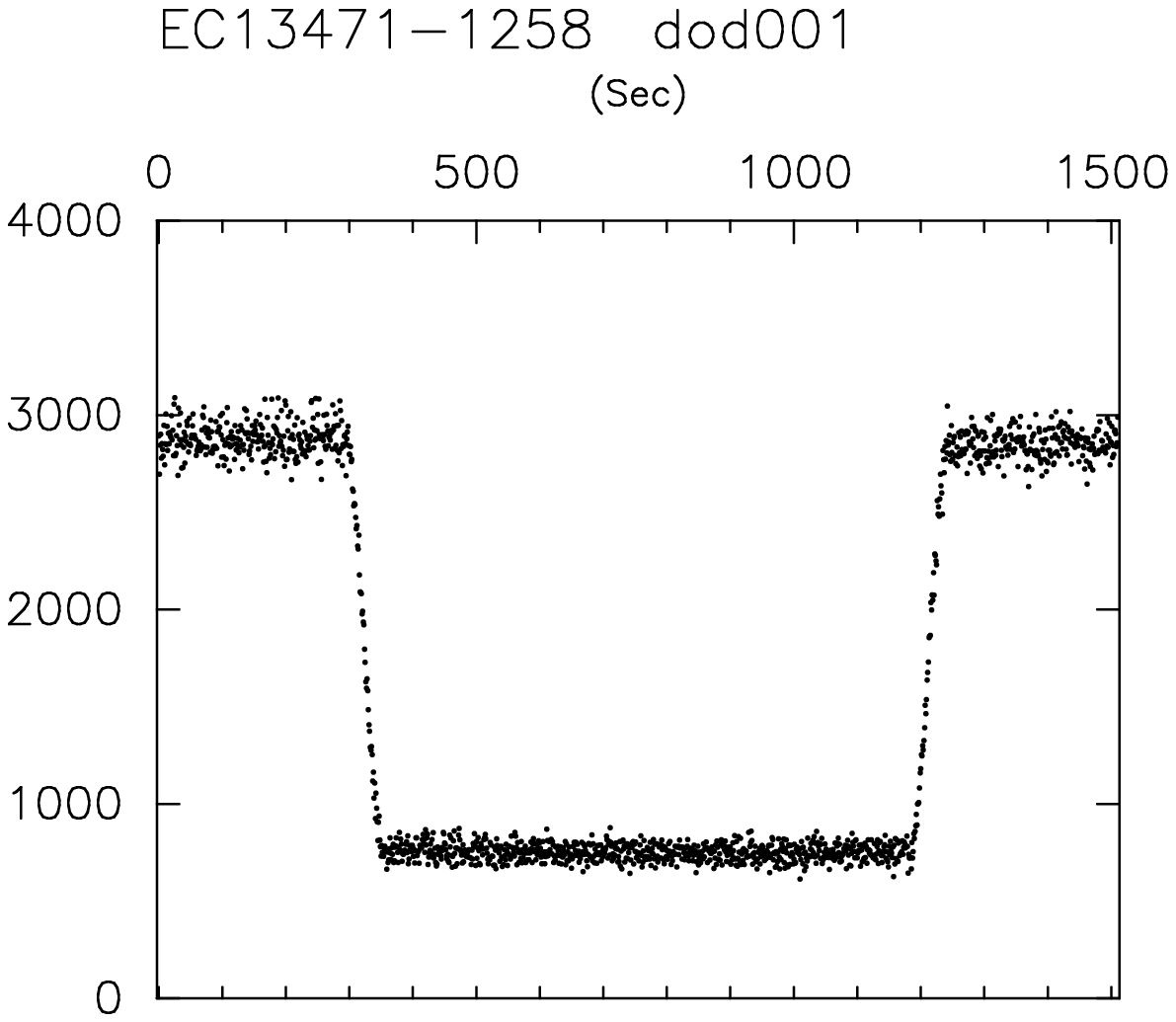,height=4cm,angle=0} 
   \end{tabular}
   \end{center}
   \caption[example]  
   { \label{fig:fig1}	   
     (Upper) $V$ band light curve at 30 s time resolution. The ordinate is
     extinction corrected counts s$^{-1}$ in the natural unfiltered, 
     photometric system. (Lower) White light eclipse light curve at 1 s time
     resolution. The ordinate is extinction corrected counts s$^{-1}$.}
   \end{figure}  

   \begin{figure} 
   \begin{center}
   \begin{tabular}{c}
   \psfig{figure=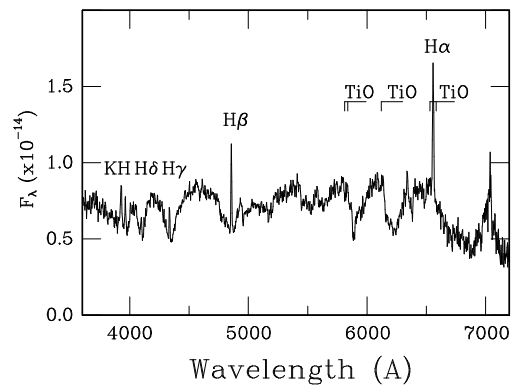,height=5.0cm,angle=0} 
   \end{tabular}
   \end{center}
   \caption[example]  
   { \label{fig:fig2}	   
   Spectrum of EC13471--1258 on 1993 Apr 15. The abscissa is in \AA\ 
   and the ordinate is in erg s$^{-1}$ cm$^{-2}$ \AA$^{-1}$. The 
   accuracy of the flux distribution shortward of 3800 \AA\ and
   longward of 6800 \AA\ is uncertain.}
   \end{figure}  

EC13471--1258 was discovered as part of the Edinburgh--Cape blue object
survey (Stobie et al. 1997, Kilkenny et al. 1997). Eclipses were
detected serendipitously by one of us (DOD) during measurement of the 
$UBV$ colours of the object: the observer could not identify why the count 
rate from the star declined suddenly, despite the apparently clear weather
and normal behaviour of the instrument. The subsequent ``unassisted" recovery 
of the count rate heralded the realization that a rapid eclipse had been 
observed and a program of high speed photometry was initiated to study the 
eclipses. These data will be considered in detail in the next section. 
Fig. 1 (upper) shows a Johnson $V$ band light curve at 30 s time 
resolution. The binary period, 0.151 d (3.62 hr), can be easily estimated 
from the horizontal axis. Also obvious is a modulation at half the orbital 
period resulting from the observer viewing at different aspect the 
geometrically distorted uneclipsed star (an ``ellipsoidal modulation").
Fig. 1 (lower) shows one of the better ``white light" eclipses at 1 s 
time resolution. The eclipse is 1.5 mag deep and total. The time from first 
to last contact is 944 s, the duration of totality is 835 s and ingress (or 
egress) take place in 54 s. The eclipses are slightly deeper in white light
than in the $V$ band.

\begin{table*}
\caption{High speed white light photometry and eclipse timings of EC13471--1258}
\begin{tabular}{@{}lcrcclccrcr}
Run &   Date &  \multicolumn{1}{c}{Start of Run}  &  Length  
             & Cycle & Mid Ingress &  Mid Egress \\
Name &       &  \multicolumn{1}{c}{JD$_{\odot}$}           &   (hr)   
     &       &  \multicolumn{1}{c}{JD$_{\odot}$}  & \multicolumn{1}{c}{JD$_{\odot}$}\\
     &       &  \multicolumn{1}{c}{2440000+}    &          
     &       &  \multicolumn{1}{c}{2440000+}    & \multicolumn{1}{c}{2440000+}\\
\\
S5465  & 06 3 1992 &  8687.553399 & 2.1 &       &               \\                                
S5469  & 07 3 1992 &  8689.460013 & 1.7 &       &               \\                                
S5470  & 08 3 1992 &  8689.570206 & 1.9 &       &               \\                                
S5473  & 10 3 1992 &  8691.585272 & 0.6 &    13 &   8691.595195 \\                                
ck0008 & 02 4 1992 &  8714.511519 & 0.3 &   165 &                &   8714.520633 \\   
ck009b & 02 4 1992 &  8715.409417 & 0.5 &   171 &   8715.414876  &   8715.425178 \\   
ck0011 & 03 4 1992 &  8715.557234 & 0.6 &   172 &   8715.565678  &   8715.575927 \\   
S5478  & 09 4 1992 &  8721.587396 & 0.5 &   212 &   8721.595947  &   8721.606231 \\   
S5480  & 11 4 1992 &  8723.551101 & 0.4 &   225 &   8723.555775  &   8723.566070 \\   
S5494  & 27 5 1992 &  8770.222061 & 2.2 &   535 &   8770.290602  &   8770.300917 \\   
S5494a & 27 5 1992 &  8770.315458 & 2.0 &       &                &               \\   
S5494b & 27 5 1992 &  8770.398786 & 1.4 &       &                &               \\   
S5498  & 28 5 1992 &  8771.338662 & 0.5 &   542 &   8771.345941  &   8771.356243 \\   
\\
S5585  & 24 2 1993 &  9043.449210 & 0.7 &  2347 &   9043.463426  &   9043.473720 \\   
S5587  & 25 2 1993 &  9043.607556 & 0.5 &  2348 &   9043.614198  &               \\   
S5591  & 26 2 1993 &  9044.509713 & 0.6 &  2354 &   9044.518725  &   9044.529037 \\   
S5597  & 27 2 1993 &  9046.463340 & 0.8 &  2367 &   9046.478580  &   9046.488874 \\   
ck29   &  194 1993 &  9097.423855 & 0.6 &  2705 &   9097.434649  &   9097.444946 \\   
S5621  & 21 6 1993 &  9160.288209 & 0.6 &  3122 &   9160.300567  &   9160.310872 \\   
\\
S5678  & 15 1 1994 &  9367.581451 & 0.6 &  4497 &   9367.592196  &   9367.602487 \\   
S5702  & 10 3 1994 &  9421.551509 & 0.6 &  4855 &   9421.563384  &   9421.573684 \\   
S5704  & 11 3 1994 &  9422.605664 & 0.6 &  4862 &                &   9422.628966 \\   
ck70   & 16 5 1994 &  9489.395598 & 2.3 &  5305 &   9489.404255 \\                                
S5754  & 10 6 1994 &  9514.406446 & 0.9 &  5471 &   9514.430010  &   9514.440315 \\   
S5760  & 12 6 1994 &  9516.226818 & 0.6 &  5483 &                &   9516.249406 \\   
\\
S5831  & 03 3 1995 &  9779.605904 & 0.5 &  7230 &   9779.612423  &   9779.622709 \\   
S5837  & 04 3 1995 &  9781.411594 & 0.3 &  7242 &   9781.421522 \\                                
S5840  & 05 3 1995 &  9782.469310 & 0.5 &  7249 &   9782.476814  &   9782.487129 \\   
S5853  & 24 5 1995 &  9862.220185 & 0.6 &  7778 &   9862.227543  &   9862.237838 \\   
S5858  & 27 5 1995 &  9865.392234 & 0.4 &  7799 &                &   9865.403793 \\   
ck120  & 31 5 1995 &  9869.405236 & 1.7 &  7826 &   9869.463916  &   9869.474208 \\   
ck122  & 01 6 1995 &  9870.215706 & 0.3 &  7831 &   9870.217693  &   9870.227990 \\   
\\
dmk024 & 28 1 1996 & 10110.515827 & 0.5 &  9425 &  10110.525155  &  10110.535446 \\   
S5865  & 22 2 1996 & 10135.542653 & 0.5 &  9591 &  10135.550900  &  10135.561204 \\   
S5868  & 23 2 1996 & 10137.495595 & 0.4 &  9604 &  10137.510741 \\                                
S5871  & 25 2 1996 & 10138.560961 & 0.4 &  9611 &  10138.566058  &  10138.576355 \\   
m0165c & 22 3 1996 & 10164.548170 & 0.4 &       &               \\                                
m0280b & 15 7 1996 & 10280.237416 & 1.3 & 10551 &  10280.278120  &  10280.288420 \\   
\\
S5892  & 14 2 1997 & 10493.585765 & 0.7 & 11966 &  10493.600012  &  10493.610315 \\   
dod001 & 06 5 1997 & 10575.305595 & 0.5 & 12508 &  10575.310609  &  10575.320907 \\   
\\
dod025 & 19 8 1998 & 11045.216518 & 0.4 & 15625 &  11045.221857  &  11045.232149 \\
\\
m1254d & 17 3 1999 & 11254.558177 & 1.9 & 17014 &  11254.623987  &  11254.634289 \\
run001 & 12 5 1999 & 11311.386676 & 2.1 & 17391 &  11311.459600  &  11311.469906 \\
\\
dod052 & 21 1 2002 & 12295.600757 & 0.4 & 23919 &  12295.604485  &  12295.614793 \\
\\
\end{tabular}
\end{table*}

The absence of a secondary eclipse indicates that the cool secondary 
is much larger than the hot primary; this is confirmed by the optical 
spectrum of EC13471--1258 which is shown in Fig. 2 at 7 \AA\ resolution. 
These data were obtained through a wide slit in order to obtain an
accurate estimate of the flux distribution; we believe that the raised 
flux below 3800 \AA\ and sagging flux above 6800 \AA\ are 
instrumental artifacts, but that the rest of the spectrum is
accurate. The flux distribution is quite flat: in the red region of the 
spectrum, the TiO molecular bands and other features typical of an M dwarf 
are visible, while the most prominent features in the blue are the broad 
Balmer absorption lines of a white dwarf. In addition, narrow emission 
cores to the Balmer lines can be seen (especially at H$\alpha$ and H$\beta$)
as well as the Ca\,II H and K lines in emission.

Figs. 1 and 2 show that EC13471--1258 is a short period eclipsing binary
comprising a white dwarf and an M dwarf. The plan of the paper is first
to present all the observational data and associated analysis. The last 
part will use the observational constraints to determine the binary
system parameters and discuss the evolutionary status of the system.
The approach will be similar to that applied to V471 Tau by O'Brien,
Bond \& Sion (2001). An earlier observational study of EC13471--1258 has
been published by Kawka et al. (2002).

\section{High speed white light photometry}

Time series photometry was obtained in ``white light" with 1 s integrations
using blue sensitive photomultiplier tubes on the SAAO 0.75 and 1.0 m
telescopes. Data were obtained during nine observing seasons spanning ten 
years. An observing log is given in Table 1.

The photomultiplier data were reduced by subtracting the sky background 
and correcting for atmospheric extinction using a mean extinction
coefficient of 0.3 mag/airmass. Four different photomultiplier tubes
were used; although all were blue sensitive, small differences in
red cutoff may affect the depth of the eclipses. This possibility was
kept in mind in the analysis presented below. Care was taken to ensure 
that the heliocentric Julian date was calculated as accurately as possible
so that the eclipse times are precisely determined.

We searched the longer light curves for evidence for rapid oscillations. No
significant peaks in the Fourier amplitude spectrum were found, with a
limit of 1 mmag.

\subsection{The ephemeris}

The best quality light curves were selected and times
of mid ingress and mid egress were measured by fitting three straight
lines to the data around ingress (or egress) by least squares. One 
line was fitted to the out-of-eclipse data (with zero slope, of course).
A second line with zero slope was fitted to the section in totality 
just after ingress (or before egress). The third straight line was 
fitted to the ingress (or egress) section. The time of mid ingress 
(or egress) was taken to be the time when the third line crossed
the average levels of the first two lines. These timings (for both
ingress and egress are collected in Table 1). Run S5470 defines
cycle zero but no timings are given. The reason for this is that
the eclipse observed in S5470 was distorted by thin cirrus. Cloud
interference usually explains the missing values associated with some
of the other runs  in Table 1 (including those where only the ingress or
egress timing is listed). 

By the third observing season it became apparent that the eclipses
could not be explained by a linear ephemeris (i.e. a constant
period). A quadratic ephemeris was thus fitted to the eclipse
timings as more data were accumulated for a few years. However, in 
the last several years this quadratic ephemeris has proved to have 
little predictive power. Figs. 3(a) and (b) show the timing residuals
for egress and ingress, respectively, with respect to linear (Fig. 3a) 
and quadratic (Fig. 3b) ephemerides.  It is clear that the scatter of 
the eclipse timings with respect to these ephemerides is far larger 
than can be attributed to measurement error (which can be estimated 
from the scatter in the groups of timings made within a few days of 
each other).

The question might be raised as to why we have not converted all the timings 
to Barycentric Julian Dynamical Date. The reason is that, as already shown, 
there are substantial residuals, far in excess of measurement error. The 
difference between Barycentric timings and Heliocentric timings is a linear 
shift in the time base (due to the addition of leap seconds in the UTC time 
system), and a periodic error of at most 2 s (with roughly the orbital period 
of Jupiter). This periodic error is quite small compared to the residuals
seen in Fig. 3(a,b) ($\sim12$ s). We thus prefer to use Heliocentric timings 
for the present and if future observations and analysis warrant it, it will be
relatively straightforward to convert the timings listed in Table 1.

   \begin{figure} 
   \begin{center}
   \begin{tabular}{c}
   \psfig{figure=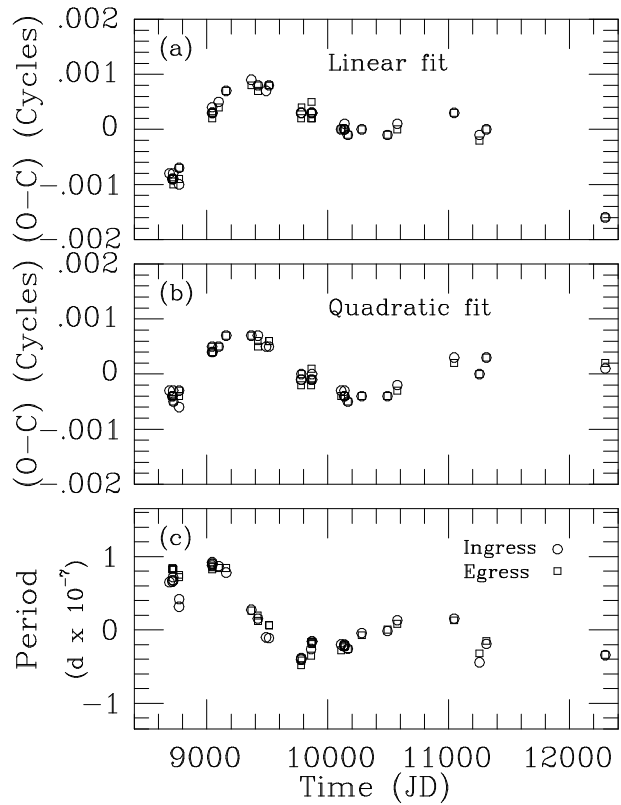,height=10.6cm,angle=0} 
   \end{tabular}
   \end{center}
   \caption[example]  
   { \label{fig:fig3}	   
   (a) Fit of a linear ephemeris to the ingress (open circles
   and egress timings (open squares). (b) Fit of a quadratic 
   ephemeris to the ingress and egress timings. (c) The mean period 
   estimated from the ingress (open circles) and egress (open squares) 
   data. The zeropoint on the vertical axis is at 0.150757525 d. The 
   points show raw estimates; an interpolated retrospectively smoothed 
   estimate passes through the points (e.g. Koen 1996, and references 
   therein). For all 3 panels the Julian Date on the abscissa is with 
   respect to 2440000.0.}
   \end{figure}  

\begin{table}
\caption{Ephemerides for EC13471--1258}
\begin{tabular}{@{}crrrr}
Parameter & \multicolumn{1}{c}{Ingress} & \multicolumn{1}{c}{Mid-Eclipse}
          & \multicolumn{1}{c}{Egress}  \\
          & \multicolumn{1}{c}{2440000+} & \multicolumn{1}{c}{2440000+}
          & \multicolumn{1}{c}{2440000+}\\
\\
Epoch     & 8689.63547 & 8689.64062 & 8689.64577\\
(JD$_{\odot}$) & $\pm$ 1 & $\pm$ 1 & $\pm$ 1 \\
\\
Period    & 0.150757525 &  0.150757525 & 0.150757525\\
 (d)      & $\pm$ 1 & $\pm$ 1 & $\pm$ 1 \\
\\
\end{tabular}
\end{table}

Although the residuals with respect to the quadratic ephemeris are smaller 
than those of the linear ephemeris (as expected from including an additional 
parameter in the model), there is little to choose between the two. 
Furthermore, we believe that there is little point in deriving a 
sophisticated functional fit to the eclipse timings in Table 1: there 
is jitter in the binary period of the system which is intrinsic and any 
functional fit now would likely have no predictive power.  Instead, 
we derive simple linear ephemerides for mid-ingress, mid-egress and 
conjunction as used in: 
\begin{eqnarray}
{\rm JD_{\odot}^{Ingress}=Epoch+(Period \times E)} \nonumber \\
{\rm JD_{\odot}^{Mid-eclipse}=Epoch+(Period \times E)} \nonumber \\
{\rm JD_{\odot}^{Egress}=Epoch+(Period \times E)} \nonumber
\end{eqnarray}
where E is the cycle number. The parameters are listed in Table 2.
Errors on the quantities are shown in Table 2: the systematic 
behaviour of the residuals precludes any formal estimate of these 
errors. Instead, these error estimates were based on the fact that
each eclipse can be timed with an accuracy of about 1 s and that the
observations span nearly 24000 cycles. There are, however, systematic 
departures of up to $\sim\pm12$ sec with respect to these ephemerides. 

Koen (1996) proposed a statistical model for sparsely observed phases of 
interest (in the present case eclipses) of periodic stars. The observed 
eclipse timings $T_j$ are assumed to be determined by three independent 
processes: (i) the slowly evolving mean period $P_j$ of the binary; (ii) 
small, random, cycle-to-cycle variations $\epsilon_j$ in the binary period; 
and (iii) measurement error $e_j$. Assuming that the cycle count difference 
between the $j$-th and $(j-1)$-th observation is $n_j$,
\begin{equation}
T_j-T_{j-1}=n_jP_j+\sum_{k=N_{j-1}}^{N_j}\epsilon_k+e_j
\equiv n_jP_j+U_j+e_j
\end{equation}
where $\epsilon_j$ and $e_j$ are zero mean white noise processes with 
variances $\sigma^2_\epsilon$ and $\sigma^2_e$, and $N_j$ is the cumulative 
cycle count (i.e. $N_j=N_{j-1}+n_j$). The specification is completed by a 
model for $P_j$: Koen(1996) showed by simulation that the random walk 
assumption
\begin{equation}
P_j=P_{j-1}+\sum_{k=N_{j-1}}^{N_j}\eta_k
\end{equation}
(where $\eta_k$ is zero mean white noise with variance $\sigma^2_\eta$) works 
very well. It follows that the variances of the three terms on the right hand 
side of (1) are
$${\rm var}(n_jP_j)=n_j^3 \sigma_\eta^2 \;\;\;\;\;\;\;\;
{\rm var}(U_j)=n_j\sigma^2_\epsilon \;\;\;\;\;\;\;\; 
{\rm var}(e_j)=\sigma^2_e$$ 

Eqns. (1) and (2) can be solved with the aid of the Kalman filtering algorithm;
see Koen (1996) for details. The optimal solution (according to the Bayes
Information Criterion -- see any modern text on time series analysis, or
Koen 1996) has $\sigma_\epsilon=0$, $\sigma_\eta=8.7E-10$ and 
$\sigma_e=1.05E-5$ (i.e. 0.9 s), based on the ingress data. The same model
is selected for the egress data; parameter values are $\sigma_\eta=7.4E-10$
and $\sigma_e=1.36E-5$ (1.2 s). The solutions for $P_j$ are plotted in Fig. 
3(c). These results show that the period variations are not due to cycle
to cycle variations but due to a random walk in the binary period. 
Although there is a general trend of period shortening in Fig. 3(c) with
substantial variations about this trend, this behaviour is applicable
only to the observation period; inferences about the future behaviour
of the orbital period cannot be drawn from these results.

   \begin{figure} 
   \begin{center}
   \begin{tabular}{c}
   \psfig{figure=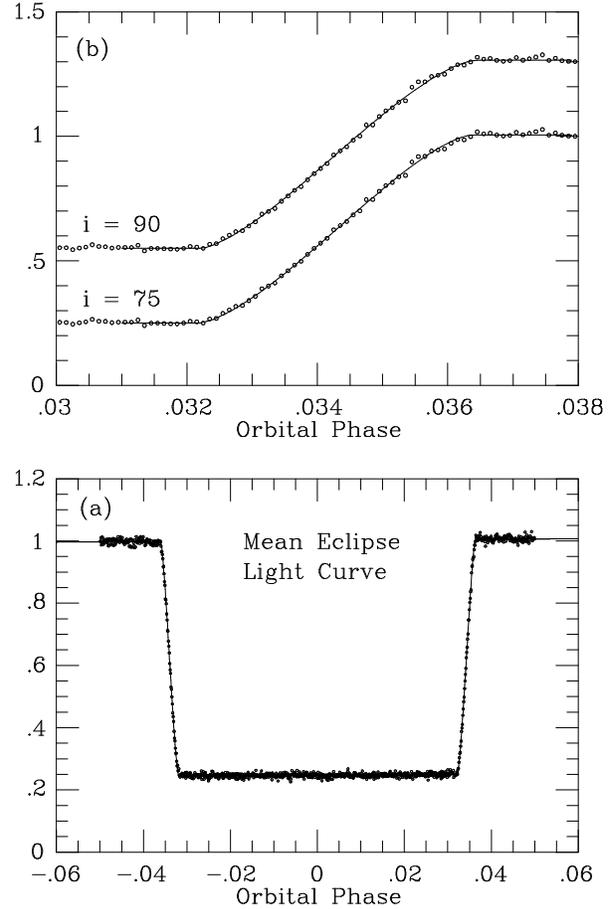,height=12.0cm,angle=0} 
   \end{tabular}
   \end{center}
   \caption[example]  
   { \label{fig:fig4}	   
   (a) Mean white light eclipse curve with a model
   superimposed as described in the text. The ordinate is in
   units of the brightness just prior to eclipse. (b) As for (a)
   except that only the egress section is plotted (on a larger scale).
   Model fits to the egress for $i=75^{\rm o}$ and $i=90^{\rm o}$ are 
   superimposed on the data. The data and the model for $i=90^{\rm o}$ 
   have been shifted vertically by 0.3 for clarity.}
   \end{figure}  

\subsection{Eclipse analysis}

The white light eclipse light curves (which are in extinction corrected
counts s$^{-1}$) were normalized by the count rate just prior to ingress
and the eclipse ephemeris was used to fold all the white light eclipses into
1000 orbital phase bins between phase -0.05 and 0.05. Small corrections were
used to the eclipse timings to reduce the intrinsic jitter and precisely
``align" the eclipses. The result is shown in Fig. 4(a). The resolution of 
0.0001 in phase corresponds to $\sim1.3$ s, comparable to the original time 
resolution of each individual eclipse. The reduction in scatter compared 
to the eclipse plotted in Fig. 1 is obvious. Analysis of the light curves 
of totally eclipsing binaries can, in principle, provide $R_1/a$, $R_2/a$ 
and $i$ where, in the present context, $R_1$ is the radius of the white dwarf, 
$R_2$ is the radius of the M dwarf, $a$ is the orbital separation and $i$ is 
the inclination (throughout this paper, subscripts 1 and 2 shall refer to the 
white dwarf and M dwarf respectively). However, Ritter \& Schroder (1979: 
RS79) have shown that because eclipses of the kind seen in Fig. 4(a) have 
only two well-defined properties, the times from first to fourth and
second to third contacts, the analysis of such eclipses cannot
provide a unique solution to all three quantities. For the sake of
brevity, we do not repeat RS79's arguments here and refer the reader to 
their paper. We do, however, follow their analysis closely and we fit 
the mean eclipse in Fig. 4(a) (solid line) with a slightly
different analytic function to that which they derived:
\begin{eqnarray*}
I(t) & = & A_0 + A_1(t-t_0) + d.f(R_1/a,R_2/a,i,t-t_0) \\
\\
 f   & = & 1,\ \ \ \ \ \ \ \ \ \ \ \ \ \ \ \ \ \ \ \ \ \ \ \ \ \ \ \ {\rm for}\  x < -\rho \\
 f   & = & 0.5 + 1/\pi\rho^2 [\rho^2{\rm sin}^{-1}(s/\rho) \\
     &   & + s(\rho^2-s^2)^{1/2} - [(\rho^2-s^2)(1-\rho^2+s^2)]^{1/2}\\
     &   & + {\rm sin}^{-1}(\rho^2-s^2)^{1/2}]\ \ \ \ \ \ \ \ \ \ \  {\rm for}\ -\rho<x<\rho\\
 f   & = & 0,\ \ \ \ \ \ \ \ \ \ \ \ \ \ \ \ \ \ \ \ \ \ \ \ \ \ \ \ {\rm for}\ x > \rho \\
\\
 s   & = & (2x-x^2-\rho^2)/(2(1-x))\\
\rho & = & R_1/R_2 \\
 x   & = & 1 - (4\pi^2/P^2(t-t_0)^2+{\rm cos^2}i)^{1/2}/(R_2/a)\\
\end{eqnarray*}
  
\noindent
In the above equations, $I(t)$ describes the brightness of the system 
near and during eclipse. $A_0$ is the brightness during eclipse and 
$d$ is the depth of the eclipse. $A_1$ allows for a linear variation in 
brightness during the observation. $f$ is a function that varies between 
0 and 1 according to the value of $x$: $x<-\rho$ outside eclipse and $x>\rho$
during totality. $P$ is the orbital period and $t_0$ is the time of mid
eclipse. The indeterminacy of these equations was resolved by choosing
a value for the inclination angle $i$. The function $I$ was then fitted
to the light curve using a nonlinear least squares technique, optimizing
6 parameters: $A_0, A_1, d, t_0, R_1/a$ and $R_2/a$. These calculations
were performed for a range of $i$ from 70$^{\rm o}$ to 90$^{\rm o}$ in
steps of 1$^{\rm o}$. The parameters $A_0, A_1, t_0$ and $d$ did not
vary with $i$. Their values were found to be: $A_0 = 0.247\pm0.001, 
d = 0.755\pm0.001$i, ; $A_1 = 0.090\pm0.001$ (in units of the out-of-eclipse 
brightness) and $t_0 = 0.0\ {\rm s}$. The positive slope, $A_1$, during 
totality is significant and will be discussed further below. The most 
interesting quantities from the eclipse analysis are the stellar radii. 
These are listed in Table 3 as a function of $i$. The model fit for
inclination 75$^o$ (an arbitrary choice) is shown as the solid line in
Fig. 4(a).

Fig. 4(b) shows the mean eclipse egress on a larger scale with the
models for $i=75^{\rm o}$ and $i=90^{\rm o}$ superimposed. Close
inspection shows that the models are in satisfactory agreement with
the data and, as pointed out by RS79, essentially identical. Even
with the high signal-to-noise in the mean eclipse curve, the data
are incapable of distinguishing between the two models. The models for
75$^o$ and 90$^o$ were divided by each other to show the subtle difference
between the models. The data were also divided by each model in turn
to see if any such trend could be discerned. Again, there was no basis 
from these operations on which to fix the inclination.

The traditional eclipse parameters: the time from first to fourth 
contact ($\phi_{\rm 14}$), the time from second to third contact 
($\phi_{\rm 23}$), and the time from mid-ingress to mid-egress of
the primary ($\phi_{\rm 1/2}$) (e.g. see equations 2-4 of Robinson,
Nather \& Patterson 1978), were determined to be:
\begin{eqnarray*}
\phi_{\rm 14} &  = & 0.07246 \pm 0.00010 \ \ \  (944 \pm 1.3 s)  \\ 
\phi_{\rm 23} &  = & 0.06408 \pm 0.00005 \ \ \  (835 \pm 1.0 s)  \\ 
\phi_{\rm 1/2} & = & 0.06832 \pm 0.00001 \ \ \  (889.9 \pm 0.2 s) 
\end{eqnarray*}

\begin{table}
\caption{Eclipse parameters}          
\begin{center}
\begin{tabular}{@{}ccc}  
 $i$  & $R_2/a$ & $R_1/a$\\
\\
  70 &  0.404 &   0.00695 \\
  71 &  0.390 &   0.00720 \\
  72 &  0.376 &   0.00746 \\
  73 &  0.363 &   0.00775 \\
  74 &  0.349 &   0.00805 \\
  75 &  0.336 &   0.00837 \\
  76 &  0.323 &   0.00870 \\
  77 &  0.311 &   0.00906 \\
  78 &  0.299 &   0.00944 \\
  79 &  0.287 &   0.00983 \\
  80 &  0.276 &   0.01023 \\
  81 &  0.266 &   0.01064 \\
  82 &  0.256 &   0.01106 \\
  83 &  0.247 &   0.01148 \\
  84 &  0.239 &   0.01187 \\
  85 &  0.232 &   0.01225 \\
  86 &  0.226 &   0.01258 \\
  87 &  0.221 &   0.01286 \\
  88 &  0.217 &   0.01307 \\
  89 &  0.215 &   0.01320 \\
  90 &  0.215 &   0.01325 \\
\\
\end{tabular}
\end{center}
\end{table}

All good quality individual runs were also fitted with the model.
No significant variations in eclipse widths were found. There was 
some variation in the eclipse depths but given the uncontrolled 
photometry (different telescopes and detectors as well as no 
observations of standard stars), these modest variations cannot 
be attributed to the star. However, the significant positive slope
seen in the model of the mean eclipse was found in many of the
individual runs (as expected as the mean eclipse was derived from
the individual runs). This positive slope is certainly not of
instrumental origin but arises from the binary. In roughly fifty per 
cent of high speed photometry runs, a positive slope (of variable size) 
during eclipse was detected; no negative slope was ever detected; in the 
remaining cases, the slope during eclipse was consistent with zero. We 
discuss this point further in the section on multicolour photometry.

   \begin{figure} 
   \begin{center}
   \begin{tabular}{c}
   \psfig{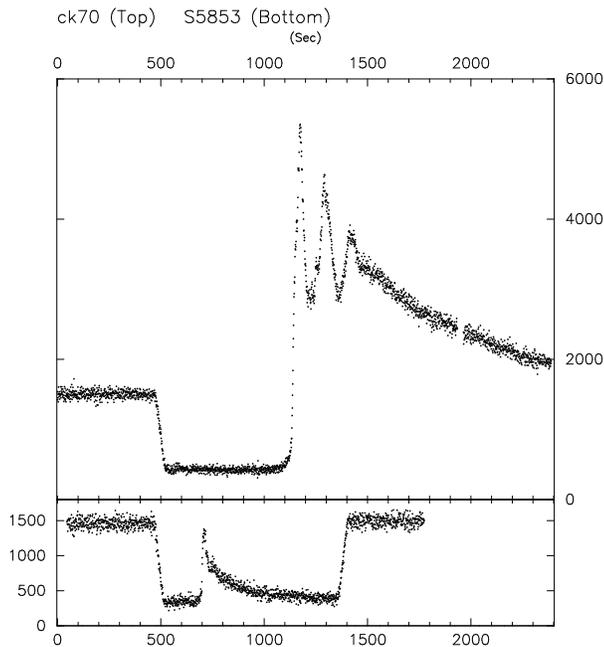} 
   \end{tabular}
   \end{center}
   \caption[example]  
   { \label{fig:fig5}	   
   (Upper) Run ck70 showing a substantial and double-peaked flare. 
   The ordinate is in extinction corrected counts s$^{-1}$. (Lower)
   Run S5853 also shows a single-peaked flare of lower amplitude.}
   \end{figure}  

\begin{table}
\caption{Photoelectric $UBV$ photometry}
\begin{tabular}{@{}cccccc}
 JD$_{\odot}$ & $V_{\rm tot}$ & $B-V_{\rm tot}$ & $U-B_{\rm tot}$ & $B_{\rm tot}$ & $U_{\rm tot}$\\
 2440000+ \\
      \multicolumn{6}{c}{1992 Mar 04/05}\\
 8686.634 & 14.42 & 0.63 & -0.54 & 15.05 & 14.51 \\
\\
      \multicolumn{6}{c}{1992 Mar 06/07}\\
 8688.544 & 14.37 & 0.63 & -0.54 & 15.00 & 14.46 \\
 8688.570 & 14.47 & 0.58 & -0.58 & 15.05 & 14.47 \\
 8688.596 & 14.43 & 0.56 & -0.54 & 15.00 & 14.45 \\
\\
      \multicolumn{6}{c}{1992 Mar 08/09}\\
 8690.545 & 15.21 & 1.54 &  0.36 & 16.75 & 17.10 \\
 8690.564 & 14.41 & 0.59 & -0.54 & 15.01 & 14.47 \\
\\
      \multicolumn{6}{c}{1992 Mar 09/10}\\
 8691.458 &  7.15$^*$ & 0.42 &  0.07 \\
 8691.473 & 14.52 & 0.59 & -0.53 & 15.11 & 14.58 \\
 8691.481 & 14.36 & 0.65 & -0.52 & 15.00 & 14.49 \\
 8691.485 & 14.36 & 0.65 & -0.52 & 15.01 & 14.49 \\
 8691.493 &  7.13$^*$ & 0.42 &  0.07 \\
 8691.497 & 14.34 & 0.65 & -0.54 & 14.99 & 14.45 \\
 8691.502 & 14.39 & 0.60 & -0.51 & 14.98 & 14.47 \\
 8691.511 & 14.45 & 0.56 & -0.55 & 15.01 & 14.47 \\
 8691.516 & 14.46 & 0.56 & -0.66 & 15.09 & 14.36 \\
 8691.524 &  7.13$^*$ & 0.42 &  0.07 \\
 8691.528 & 14.49 & 0.55 & -0.60 & 15.04 & 14.44 \\
 8691.534 & 14.48 & 0.55 & -0.57 & 15.03 & 14.45 \\
 8691.539 & 14.44 & 0.57 & -0.57 & 15.01 & 14.44 \\
\\
      \multicolumn{6}{c}{1992 May 29/30}\\
 8772.222 &  7.15$^*$ & 0.40 &  0.06 \\
 8772.229 & 14.45 & 0.58 & -0.52 & 15.03 & 14.50 \\
 8772.235 & 14.42 & 0.60 & -0.52 & 15.02 & 14.51 \\
 8772.241 & 14.47 & 0.58 & -0.49 & 15.05 & 14.55 \\
 8772.271 &  7.14$^*$ & 0.41 &  0.05 \\
 8772.282 & 14.42 & 0.59 & -0.50 & 15.01 & 14.51 \\
 8772.289 & 14.38 & 0.62 & -0.53 & 15.00 & 14.47 \\
 8772.295 & 14.39 & 0.60 & -0.53 & 14.98 & 14.45 \\
 8772.301 &  7.15$^*$ & 0.41 &  0.06 \\
 8772.304 & 14.43 & 0.57 & -0.54 & 15.00 & 14.47 \\
 8772.310 & 14.44 & 0.59 & -0.56 & 15.03 & 14.47 \\
 8772.316 & 14.45 & 0.59 & -0.56 & 15.03 & 14.47 \\
 8772.321 &  7.15$^*$ & 0.41 &  0.06 \\
 8772.325 & 14.48 & 0.47 & -0.68 & 14.95 & 14.27 \\
 8772.331 & 14.48 & 0.53 & -0.63 & 15.02 & 14.38 \\
 8772.337 & 14.50 & 0.52 & -0.60 & 15.03 & 14.43 \\
 8772.343 &  7.15$^*$ & 0.41 &  0.06 \\
 8772.346 & 14.46 & 0.58 & -0.60 & 15.04 & 14.44 \\
 8772.352 & 14.46 & 0.56 & -0.54 & 15.02 & 14.48 \\
 8772.358 & 14.44 & 0.57 & -0.56 & 15.01 & 14.45 \\
 8772.364 &  7.15$^*$ & 0.41 &  0.07 \\
 8772.367 & 14.39 & 0.62 & -0.51 & 15.01 & 14.51 \\
 8772.374 & 14.46 & 0.57 & -0.49 & 15.03 & 14.54 \\
 8772.380 & 14.45 & 0.61 & -0.51 & 15.06 & 14.55 \\
 8772.385 &  7.18$^*$ & 0.42 &  0.06 \\
 8772.388 & 14.51 & 0.58 & -0.57 & 15.08 & 14.52 \\
 8772.394 & 14.54 & 0.53 & -0.51 & 15.07 & 14.56 \\
 8772.421 & 14.50 & 0.58 & -0.53 & 15.08 & 14.55 \\
 8772.426 &  7.17$^*$ & 0.42 &  0.08 \\
 8772.429 & 14.45 & 0.58 & -0.52 & 15.03 & 14.51 \\
 8772.435 & 14.40 & 0.63 & -0.51 & 15.04 & 14.53 \\
 8772.440 & 14.40 & 0.61 & -0.54 & 15.01 & 14.47 \\
 8772.445 &  7.17$^*$ & 0.42 &  0.07 \\
 8772.449 & 14.39 & 0.61 & -0.53 & 15.00 & 14.46 \\
 8772.454 & 14.45 & 0.57 & -0.51 & 15.02 & 14.51 \\
 8772.460 & 14.42 & 0.64 & -0.58 & 15.06 & 14.48 \\
 8772.466 &  7.19$^*$ & 0.42 &  0.08 \\\
\end{tabular}
\end{table}

\begin{table}
\contcaption{ }
\begin{tabular}{@{}cccccc}
 JD$_{\odot}$ & $V_{\rm tot}$ & $B-V_{\rm tot}$ & $U-B_{\rm tot}$ & $B_{\rm tot}$ & $U_{\rm tot}$\\
 2440000+ \\
      \multicolumn{6}{c}{1997 Apr 29/30}\\
10568.358 & 14.39 & 0.63 & -0.55 & 14.99 & 14.44 \\
10568.366 & 14.40 & 0.63 & -0.51 & 15.03 & 14.51 \\
10568.371 & 14.43 & 0.60 & -0.53 & 15.03 & 14.50 \\
10568.379 & 15.24 & 1.55 &  0.88 & 16.79 & 17.67 \\
10568.384 & 15.22 & 1.52 &  0.70 & 16.74 & 17.43 \\
10568.389 & 14.43 & 0.59 & -0.53 & 15.02 & 14.49 \\
10568.394 & 14.40 & 0.64 & -0.53 & 15.04 & 14.52 \\
\\
      \multicolumn{6}{c}{1997 May 03/04}\\
10572.285 & 14.39 & 0.65 & -0.52 & 15.04 & 14.52 \\
10572.290 & 14.39 & 0.63 & -0.53 & 15.03 & 14.49 \\
10572.298 & 15.22 & 1.56 &  0.98 & 16.78 & 17.76 \\
10572.302 & 15.21 & 1.61 &  0.90 & 16.82 & 17.72 \\
10572.309 & 14.42 & 0.63 & -0.51 & 15.05 & 14.54 \\
10572.313 & 14.39 & 0.65 & -0.50 & 15.04 & 14.54 \\
10572.320 & 14.38 & 0.65 & -0.49 & 15.02 & 14.54 \\
10572.324 & 14.39 & 0.66 & -0.52 & 15.04 & 14.52 \\
     & {\bf 14.43} & {\bf 0.59} & {\bf -0.54} & {\bf 15.03} & {\bf 14.47} \\
\\
\\
$V_{\rm WD}$ & $B-V_{\rm WD}$ & $U-B_{\rm WD}$ & $B_{\rm WD}$ & $U_{\rm WD}$ \\
 8690.545 & 15.15 & 0.09 & -0.68 & 15.24 & 14.56 \\
\\
10568.379 & 15.13 & 0.13 & -0.70 & 15.26 & 14.56 \\
10568.384 & 15.15 & 0.12 & -0.70 & 15.27 & 14.57 \\
\\
10572.298 & 15.11 & 0.17 & -0.71 & 15.28 & 14.57 \\
10572.302 & 15.12 & 0.15 & -0.70 & 15.27 & 14.57 \\
  & {\bf 15.13} & {\bf 0.13} & {\bf -0.70} & {\bf 15.26} & {\bf 14.57} \\
\\
\\
 & $V_{\rm Sec}$ & $B-V_{\rm Sec}$ & $U-B_{\rm Sec}$ & $B_{\rm Sec}$ & $U_{\rm Sec}$ \\
 8690.545 & 15.21 & 1.54 &  0.36 : & 16.75 & 17.10 :\\
\\
10568.379 & 15.24 & 1.55 &  0.88 : & 16.79 & 17.67 :\\
10568.384 & 15.22 & 1.52 &  0.70 : & 16.74 & 17.43 :\\
\\
10572.298 & 15.22 & 1.56 &  0.98 : & 16.78 & 17.76 :\\
10572.302 & 15.21 & 1.61 &  0.90 : & 16.82 & 17.72 :\\
    & {\bf 15.22} & {\bf 1.56} & {\bf 0.76 :} & {\bf 16.78} & {\bf 17.54 :} \\
\end{tabular}
\end{table}

\begin{table}
\begin{center}
\hspace{10mm}\caption{Single filter time series photometry}
\begin{tabular}{@{}rlcccc}
\hspace{10mm} & Run &  Date &  \multicolumn{1}{c}{Start of Run} & Length & Filter \\
\hspace{10mm} & Name &      &  \multicolumn{1}{c}{JD$_{\odot}$} &  (hr)  \\
\hspace{10mm} &      &      &  \multicolumn{1}{c}{2440000+}   \\        
\\
\hspace{10mm} & S5502  & 29 5 92 & 8772.245553 & 0.5  & $V$ \\
\hspace{10mm} & S5503  & 29 5 92 & 8772.398321 & 0.5  & $U$ \\
\\
\hspace{10mm} & S5601  & 28 2 93 & 9047.481823 & 3.8  & $V$ \\
\hspace{10mm} & S5603  & 01 3 93 & 9048.433285 & 0.5  & $B$ \\
\hspace{10mm} & S5603a & 01 3 93 & 9048.454120 & 4.4  & $V$ \\
\hspace{10mm} & ck22   & 14 4 93 & 9092.426764 & 3.8  & $V$ \\  
\hspace{10mm} & ck25   & 17 4 93 & 9095.414611 & 2.4  & $V$ \\  
\hspace{10mm} & ck26   & 18 4 93 & 9096.339396 & 5.4  & $V$ \\  
\\
\hspace{10mm} & m9460d & 17 4 94 & 9460.406300 & 4.1  & $V$ \\  
\hspace{10mm} & S5758  & 11 6 94 & 9515.325139 & 3.2  & $V$ \\  
\end{tabular}
\end{center}
\end{table}

\subsection{Flares}

The ellipsoidal modulation shown by the companion M dwarf (Fig.1) indicates
that it is tidally locked to the binary period and that therefore it is
certainly rotating very rapidly. However, for M dwarfs (and later types), it
is not at all clear that chromospheric activity is correlated with rotation
(see, for example, the recent review by Hawley, Reid \& Gizis 2000). 
The presence of H$\alpha$ emission - usually regarded as the primary 
indicator of magnetic activity - is not affected by heating 
from the very close white dwarf (see later). Nevertheless, if we assume
that the M dwarf is similar to GJ 486 (as indicated by the colour comparison 
given in section 3.2 and later discussion), then the implied spectral type 
of M3.5 means that the M dwarf is completely convective and in a regime 
where at least 20 per cent of all stars show H$\alpha$ emission and an even 
greater fraction exhibit flaring activity over a large range of energies 
(see Hawley, Reid \& Gizis 2000; especially their Figs.2 and 3). The 
situation with the cool star in EC~13471--1258 may be complicated by the 
fact that it will have certainly passed through a ``common envelope'' 
configuration and may have undergone substantial mass loss through Roche 
lobe overflow (see the later discussion). Nonetheless, it might still be 
expected to find flares of the kind seen in chromospherically active M dwarfs.

Unequivocal evidence for flares is seen in the two light curves (ck70 and
S5853) plotted in Fig. 5. Both flares began in eclipse, thus identifying 
the red star as their origin. The flare in ck70 is $\sim 3$ mag 
brighter at maximum and the profile is double peaked (the third peak
is the egress of the white dwarf). The flare in S5853 is only 1.5
mag in size. The ck70 light curve continued beyond what is plotted in
Fig. 5 with the brightness of the system gradually returning to the
pre-eclipse brightness level. The rise time of the lower amplitude flare
is a few seconds whereas in ck70 it was about a minute.

We have observed two flares out of a total of $\sim40$ hours of high speed
photometry. Two additional flares were observed in multicolour photometry
(see next section) covering 21 hours. Flaring therefore occurs with a mean 
rate of 0.067 per hour. We have also seen evidence for flares in spectroscopy
(as did Kawka et al. 2002), although it is hard to quantify the rate.  

\section{Multicolour photometry}

We now consider multicolour photometric data in order to learn about 
variations due to the M dwarf and the relative contributions of each 
component star at different wavelengths.

\subsection{Photoelectric photometry}

In order to assist any future modelling of the binary, we list in Table 4 
all the photoelectric $UBV$ measurements. They were made with blue sensitive
photomultiplier tubes. The data were reduced by correcting for atmospheric
extinction and transforming to the standard system using observations of
several tens of E region standard stars (Menzies et al. 1989).
Standard stars were also observed regularly during the nights listed in 
Table 4. In addition, on the nights of 1992 Mar 9/10 and May 29/30, a  
$\sim7$th mag local comparison star HD120543, about 10 arcmin distant from 
EC13471--1258, was also observed. Its magnitude and colours are shown by 
the data in Table 4 with a superscript asterisk ($^*$) alongside. These
data are included so the reader can judge the quality of the photometry.
Note that towards the end of the night 1992 May 29/30 the comparison star
faded by a few per cent. The EC13471--1258 data were corrected by these
differences, on the assumption that the comparison star is constant. Even
if this is not so, the corrections were at most 0.03 mag in $V$ and no
correction at all in the colours. We estimate the uncertainty on the 
photometry is $\pm0.02$ mag.

Almost all the data in Table 4 are labelled with the subscript ``tot" to
indicate the sum of the light from both component stars. Five data 
points were obtained during total eclipse of the white dwarf, allowing
a separation of the contributions of each star. These separate
contributions are shown at the bottom of Table 4 with subscripted labels
``WD" and ``Sec". Bold numbers in Table 4 are means of the respective
quantities. In the case of the ``tot" data, the eclipse points were
excluded. The eclipse depths in $V$, $B$ and $U$ are 56, 81 and 93 per
cent respectively. Only 7 per cent of the $U$ light originates from the
M dwarf, so its $U$ magnitude and $U-B$ colour are not well determined; this
is denoted in Table 4 by colons.

   \begin{figure} 
   \begin{center}
   \begin{tabular}{c}
   \psfig{figure=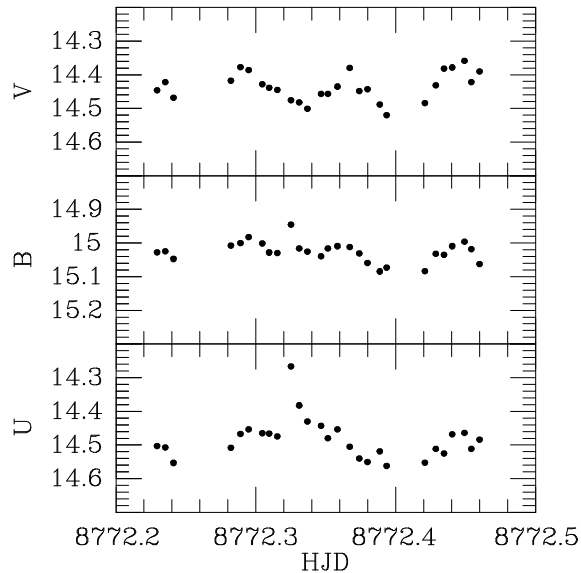,height=7.5cm,angle=0} 
   \end{tabular}
   \end{center}
   \caption[example]  
   { \label{fig:fig6}	   
   $UBV$ photometry on 29/30 May 1992 (JD 2448772). Note the
   flare in the $U$ and $B$ light curves.}
   \end{figure}  

   \begin{figure} 
   \begin{center}
   \begin{tabular}{c}
   \psfig{figure=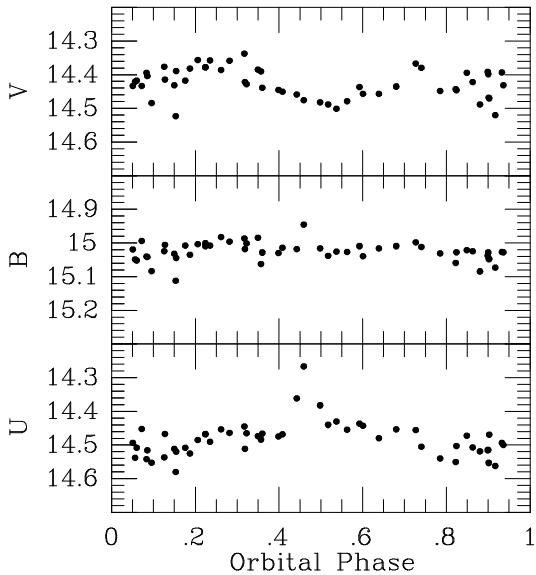,height=7.5cm,angle=0} 
   \end{tabular}
   \end{center}
   \caption[example]  
   { \label{fig:fig7}	   
   All $UBV$ photometry listed in Table 4 folded on the orbital period. 
   Eclipse points have been excluded.} 
   \end{figure}  

Fig. 6 shows the data obtained on 1992 May 29/30 (JD 2448772) excluding
the eclipses which were observed in the single filter runs S5502 and S5503 
(Table 5). The $V$ light curve shows the ellipsoidal variation (double 
humped structure) already pointed out in Fig. 1 (upper). In this case,
however, the variation is more erratic (that this is real can be judged
from the photometry of the comparison star: Table 4). The ellipsoidal
variation is less obvious in the $B$ and $U$ light curves. Instead, the $U$ 
light curve shows a flare which is less evident in $B$ and undetected at $V$. 
Such behaviour is typical of flare stars where increases in optical light 
are most obvious due to the Balmer continuum and lines going into emission.
Fig. 7 shows all the $UBV$ data in Table 4 folded on the orbital period.
As in Fig. 6, the ellipsoidal variation is obvious only in the $V$ light 
curve. The $U$ band light curve seems to reach a maximum at phase 0.5 
(even after allowing for the occurrence of a flare close to this phase).
The apparent increase in scatter in all the magnitudes close to eclipse 
is real: the effect is subtle but can be best appreciated by comparing 
the light curves at orbital phase 0.5 with phases close to 0.

Table 5 lists single filter observations made on the SAAO 0.75-m and 1-m 
telescopes with photomultiplier detectors. With the exception of the first
two runs, S5502 and S5503, which used 1 s integrations, all the other data 
were obtained with 10 s integrations. The photomultiplier data were 
reduced in the same manner as the white light data except that coefficients 
of 0.30 and 0.54 mag/airmass were used for atmospheric extinction correction 
of the $B$ and $U$ data respectively (these are close to the standard 
extinction coefficients for the site).

   \begin{figure}
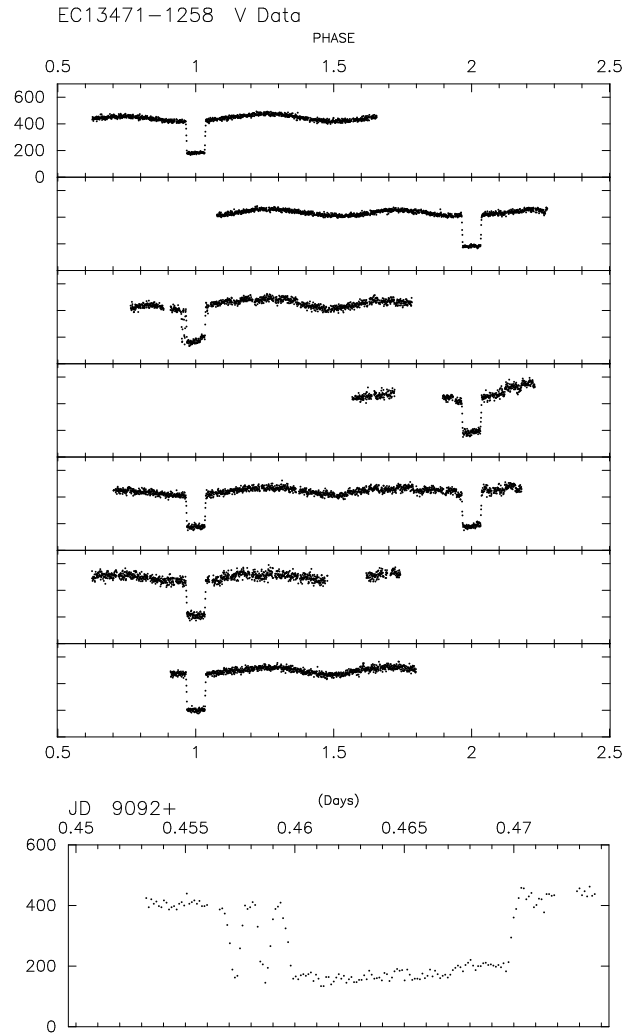
 
   \begin{center}
   \begin{tabular}{c}
   \psfig{figure=fig8a.ps,height=10.0cm,angle=0} \\
\\
   \psfig{figure=fig8b.ps,height=3.2cm,angle=0} 
   \end{tabular}
   \end{center}
   \caption[example]  
   { \label{fig:fig8}	   
   (Upper) V filter time series photometry folded on the orbital period.
   The runs are listed in Table 5 and are from the top: S5601, S5603a,
   ck22, ck25, ck26, m9460d and S5758. (Lower) Run ck22 close to eclipse
   and plotted on a larger scale. Note the two pre-eclipse dips.} 
   \end{figure}  

\begin{table*}
\caption{CCD $VRI$ photometry of EC13471--1258}
\begin{tabular}{@{}cccccccccccc}
JD$_{\odot}$ & $V$ & JD$_{\odot}$ & $R$ & JD$_{\odot}$ & $I$ & JD$_{\odot}$ & $V$ & JD$_{\odot}$ & $R$ & JD$_{\odot}$ & $I$ \\
2440000+ & & 2440000+ & & 2440000+ & & 2440000+ & & 2440000+ & & 2440000+ \\  
\\
9161.351 & 14.38 & 9161.352 & 13.67 & 9161.353 & 12.24 & 9163.268 & 14.31 & 9163.273 & 13.57 & 9163.277 & 12.11 \\
9161.354 & 14.38 & 9161.355 & 13.66 & 9161.359 & 12.32 & 9163.271 & 14.29 & 9163.276 & 13.55 & 9163.281 & 12.10 \\
9161.357 & 15.20 & 9161.358 & 13.99 & 9161.363 & 12.31 & 9163.275 & 14.28 & 9163.280 & 13.55 & 9163.284 & 12.10 \\
9161.361 & 15.22 & 9161.362 & 13.98 & 9161.366 & 12.28 & 9163.278 & 14.27 & 9163.283 & 13.54 & 9163.288 & 12.12 \\
9161.364 & 15.17 & 9161.365 & 13.97 & 9161.378 & 12.16 & 9163.282 & 14.26 & 9163.287 & 13.56 & 9163.291 & 12.13 \\
9161.367 & 14.36 & 9161.377 & 13.58 & 9161.381 & 12.15 & 9163.286 & 14.27 & 9163.290 & 13.57 & 9163.297 & 12.16 \\
9161.375 & 14.32 & 9161.380 & 13.56 & 9161.384 & 12.12 & 9163.289 & 14.29 & 9163.295 & 13.59 & 9163.300 & 12.19 \\
9161.379 & 14.32 & 9161.383 & 13.54 & 9161.388 & 12.09 & 9163.294 & 14.29 & 9163.299 & 13.63 & 9163.304 & 12.21 \\
9161.382 & 14.29 & 9161.387 & 13.51 & 9161.391 & 12.08 & 9163.298 & 14.33 & 9163.303 & 13.65 & 9163.307 & 12.23 \\
9161.385 & 14.28 & 9161.390 & 13.49 & 9161.394 & 12.06 & 9163.301 & 14.33 & 9163.306 & 13.67 & 9163.311 & 12.24 \\
9161.389 & 14.28 & 9161.393 & 13.47 & 9161.398 & 12.05 & 9163.305 & 14.36 & 9163.310 & 13.68 & 9163.314 & 12.25 \\
9161.392 & 14.25 & 9161.397 & 13.46 & 9161.401 & 12.04 & 9163.308 & 14.36 & 9163.313 & 13.68 & 9163.318 & 12.34 \\
9161.396 & 14.25 & 9161.400 & 13.46 & 9161.404 & 12.06 & 9163.312 & 14.37 & 9163.317 & 14.00 & 9163.321 & 12.33 \\
9161.399 & 14.25 & 9161.403 & 13.47 & 9161.408 & 12.09 & 9163.315 & 14.45 & 9163.320 & 13.99 & 9163.325 & 12.33 \\
9161.402 & 14.24 & 9161.407 & 13.47 & 9161.411 & 12.11 & 9163.319 & 15.17 & 9163.324 & 13.98 & 9163.328 & 12.25 \\
9161.406 & 14.26 & 9161.410 & 13.51 & 9161.415 & 12.13 & 9163.322 & 15.15 & 9163.327 & 13.65 & 9163.332 & 12.23 \\
9161.409 & 14.27 & 9161.414 & 13.53 & 9162.339 & 12.33 & 9163.326 & 14.80 & 9163.331 & 13.63 & 9163.335 & 12.20 \\
9161.412 & 14.25 & 9162.338 & 13.73 & 9162.343 & 12.33 & 9163.329 & 14.33 & 9163.334 & 13.60 & 9163.339 & 12.17 \\
9162.337 & 14.41 & 9162.342 & 13.73 & 9162.346 & 12.33 & 9163.333 & 14.32 & 9163.338 & 13.57 & 9163.343 & 12.14 \\
9162.340 & 14.42 & 9162.345 & 13.73 & 9162.349 & 12.31 & 9163.336 & 14.30 & 9163.342 & 13.55 & 9163.347 & 12.11 \\
9162.344 & 14.40 & 9162.348 & 13.72 & 9162.353 & 12.27 & 9163.341 & 14.28 & 9163.346 & 13.51 & 9163.351 & 12.09 \\
9162.347 & 14.39 & 9162.352 & 13.69 & 9162.356 & 12.24 & 9163.345 & 14.26 & 9163.350 & 13.49 & 9163.354 & 12.07 \\
9162.350 & 14.38 & 9162.355 & 13.67 & 9162.361 & 12.19 & 9163.348 & 14.25 & 9163.353 & 13.48 & 9163.358 & 12.07 \\
9162.354 & 14.37 & 9162.360 & 13.63 & 9162.365 & 12.16 & 9163.352 & 14.23 & 9163.357 & 13.47 & 9163.361 & 12.06 \\
9162.359 & 14.34 & 9162.364 & 13.59 & 9162.368 & 12.14 & 9163.355 & 14.23 & 9163.360 & 13.46 & 9163.365 & 12.09 \\
9162.362 & 14.31 & 9162.367 & 13.57 & 9162.371 & 12.13 & 9163.359 & 14.21 & 9163.364 & 13.47 & 9163.368 & 12.10 \\
9162.366 & 14.30 & 9162.370 & 13.55 & 9162.375 & 12.12 & 9163.362 & 14.22 & 9163.367 & 13.48 & 9163.372 & 12.13 \\
9162.369 & 14.29 & 9162.374 & 13.55 & 9162.378 & 12.10 & 9163.366 & 14.20 & 9163.371 & 13.52 & 9163.375 & 12.16 \\
9162.373 & 14.28 & 9162.377 & 13.54 & 9162.382 & 12.11 & 9163.369 & 14.24 & 9163.374 & 13.53 & 9163.379 & 12.19 \\
9162.376 & 14.28 & 9162.380 & 13.53 & 9162.385 & 12.13 & 9163.373 & 14.23 & 9163.378 & 13.56 & 9163.382 & 12.24 \\
9162.379 & 14.27 & 9162.384 & 13.55 & 9162.388 & 12.14 & 9163.377 & 14.27 & 9163.381 & 13.61 & 9163.386 & 12.26 \\
9162.383 & 14.28 & 9162.387 & 13.59 & 9162.392 & 12.17 & 9163.380 & 14.31 & 9163.385 & 13.65 & 9163.390 & 12.30 \\
9162.386 & 14.30 & 9162.391 & 13.60 & 9162.395 & 12.20 & 9163.384 & 14.32 & 9163.389 & 13.68 & 9163.394 & 12.29 \\
9162.389 & 14.32 & 9162.394 & 13.63 & 9162.398 & 12.22 & 9163.388 & 14.36 & 9163.393 & 13.70 & 9163.398 & 12.31 \\
9162.393 & 14.33 & 9162.397 & 13.66 & 9162.403 & 12.25 & 9163.392 & 14.36 & 9163.396 & 13.71 & 9163.401 & 12.32 \\
9162.396 & 14.35 & 9162.402 & 13.69 & 9162.406 & 12.26 & 9163.395 & 14.37 & 9163.400 & 13.71 & 9163.405 & 12.29 \\
9162.400 & 14.36 & 9162.405 & 13.69 & 9162.409 & 12.27 & 9163.399 & 14.37 & 9163.404 & 13.69 & 9163.408 & 12.27 \\
9162.404 & 14.39 & 9162.408 & 13.69 & 9162.415 & 12.35 & 9163.402 & 14.38 & 9163.407 & 13.68 & 9163.412 & 12.24 \\
9162.407 & 14.38 & 9162.414 & 14.03 & 9162.419 & 12.36 & 9163.406 & 14.37 & 9163.411 & 13.65 & 9163.415 & 12.21 \\
9162.413 & 15.22 & 9162.418 & 14.01 & 9162.422 & 12.28 & 9163.409 & 14.35 & 9163.414 & 13.62 & 9163.419 & 12.15 \\
9162.416 & 15.17 & 9162.421 & 13.97 & 9162.425 & 12.25 & 9163.413 & 14.33 & 9163.418 & 13.58 & 9163.422 & 12.14 \\
9162.420 & 15.18 & 9162.424 & 13.67 & 9162.429 & 12.25 & 9163.416 & 14.31 & 9163.421 & 13.56 & 9163.426 & 12.10 \\
9162.423 & 14.36 & 9162.428 & 13.67 & 9162.432 & 12.21 & 9163.420 & 14.28 & 9163.425 & 13.55 & 9163.429 & 12.09 \\
9162.427 & 14.34 & 9162.431 & 13.65 & 9162.436 & 12.20 & 9163.423 & 14.28 & 9163.428 & 13.53 & 9163.433 & 12.10 \\
9162.430 & 14.33 & 9162.435 & 13.61 & 9162.439 & 12.16 & 9163.427 & 14.27 & 9163.432 & 13.52 & 9164.210 & 12.23 \\
9162.433 & 14.32 & 9162.438 & 13.59 & 9162.442 & 12.15 & 9163.430 & 14.28 & 9164.209 & 13.66 & 9164.214 & 12.24 \\
9162.437 & 14.30 & 9162.441 & 13.59 & 9163.216 & 11.98 & 9164.208 & 14.35 & 9164.213 & 13.67 & 9164.217 & 12.25 \\
9162.440 & 14.28 & 9163.215 & 13.15 & 9163.220 & 12.02 & 9164.211 & 14.37 & 9164.216 & 13.69 & 9164.221 & 12.33 \\
9163.214 & 13.79 & 9163.219 & 13.23 & 9163.224 & 12.07 & 9164.215 & 14.38 & 9164.220 & 13.70 & 9164.224 & 12.34 \\
9163.218 & 13.87 & 9163.222 & 13.30 & 9163.227 & 12.12 & 9164.218 & 14.37 & 9164.223 & 14.02 & 9164.228 & 12.34 \\
9163.221 & 13.94 & 9163.226 & 13.37 & 9163.231 & 12.17 & 9164.222 & 15.22 & 9164.227 & 14.01 & 9164.231 & 12.24 \\
9163.225 & 14.01 & 9163.230 & 13.44 & 9163.234 & 12.21 & 9164.225 & 15.22 & 9164.230 & 13.94 & 9164.235 & 12.22 \\
9163.228 & 14.08 & 9163.233 & 13.49 & 9163.238 & 12.25 & 9164.229 & 15.18 & 9164.234 & 13.66 & 9164.238 & 12.21 \\
9163.232 & 14.15 & 9163.237 & 13.57 & 9163.241 & 12.28 & 9164.232 & 14.36 & 9164.237 & 13.64 & 9164.242 & 12.19 \\
9163.235 & 14.20 & 9163.240 & 13.61 & 9163.245 & 12.30 & 9164.236 & 14.35 & 9164.241 & 13.62 & 9164.245 & 12.17 \\
9163.239 & 14.25 & 9163.244 & 13.65 & 9163.248 & 12.29 & 9164.240 & 14.33 & 9164.244 & 13.60 & 9164.384 & 12.26 \\
9163.242 & 14.28 & 9163.247 & 13.67 & 9163.252 & 12.29 & 9164.243 & 14.32 & 9164.383 & 13.66 & 9164.388 & 12.23 \\
9163.246 & 14.33 & 9163.251 & 13.68 & 9163.256 & 12.27 & 9164.382 & 14.50 & 9164.387 & 13.64 & 9164.391 & 12.21 \\
9163.249 & 14.34 & 9163.255 & 13.67 & 9163.260 & 12.24 & 9164.385 & 14.34 & 9164.390 & 13.61 & 9164.395 & 12.18 \\
9163.254 & 14.34 & 9163.259 & 13.67 & 9163.263 & 12.22 & 9164.389 & 14.33 & 9164.394 & 13.59 & 9164.398 & 12.15 \\
9163.257 & 14.36 & 9163.262 & 13.64 & 9163.267 & 12.19 & 9164.393 & 14.34 & 9164.397 & 13.57 & 9164.402 & 12.13 \\
9163.261 & 14.33 & 9163.266 & 13.62 & 9163.270 & 12.15 & 9164.396 & 14.30 & 9164.401 & 13.54 & 9164.405 & 12.10 \\
9163.264 & 14.32 & 9163.269 & 13.60 & 9163.274 & 12.13 & 9164.400 & 14.31 & 9164.404 & 13.53 \\
\end{tabular}
\end{table*}

Fig. 8 (upper) shows the $V$ data from Table 5. A number of features
are apparent:

\begin{itemize}
\item Taking into account that 56 per cent of the light in the $V$ band 
      data is due to the white dwarf, the ellipsoidal variation is of 
      large amplitude: a sinusoid with period equal to half the orbital 
      period was fitted to the data in Fig. 8. The mean amplitude (peak to
      peak) of the ellipsoidal variation was found to be 22 per cent of 
      the mean brightness of the M dwarf.
\item The minimum of the ellipsoidal variation is not centred on eclipse.
      This is most noticeable in runs ck22 and ck25 (3rd and 4th from the
      top). This gives rise to the positive slope during total eclipse
      already discussed in the white light photometry. The minimum closest 
      to eclipse in the least squares fit mentioned above occurred at 
      orbital phase $-0.023 \pm 0.001$.
\item Although of low amplitude, there appears to be rapid, erratic
      variations superimposed on the overall ellipsoidal modulation.
      This is most apparent again in ck22 and ck25. For example, in
      ck22 (third panel from top) at orbital phase $\sim1.2$ there are
      rapid short time scale variations which we believe are not due
      to atmospheric transparency variations or of instrumental origin.
\item ck22 shows two pre-eclipse dips, plotted in Fig. 8 (lower).
      We are convinced that these are real and not of instrumental
      origin: the dip minima are at the level of the eclipse and not
      the sky background so the effect cannot be explained by the star
      going out of the photometer's aperture. Other tests were done
      by two of us (CK and DOD) during this observing run to search for
      any possible atmospheric or instrumental origin: no evidence was
      found. We accept, however, that there was no comparison star
      observations to {\em prove} the reality of these dips and they
      were observed only in ck22.
\end{itemize}

   \begin{figure} 
   \begin{center}
   \begin{tabular}{c}
   \psfig{figure=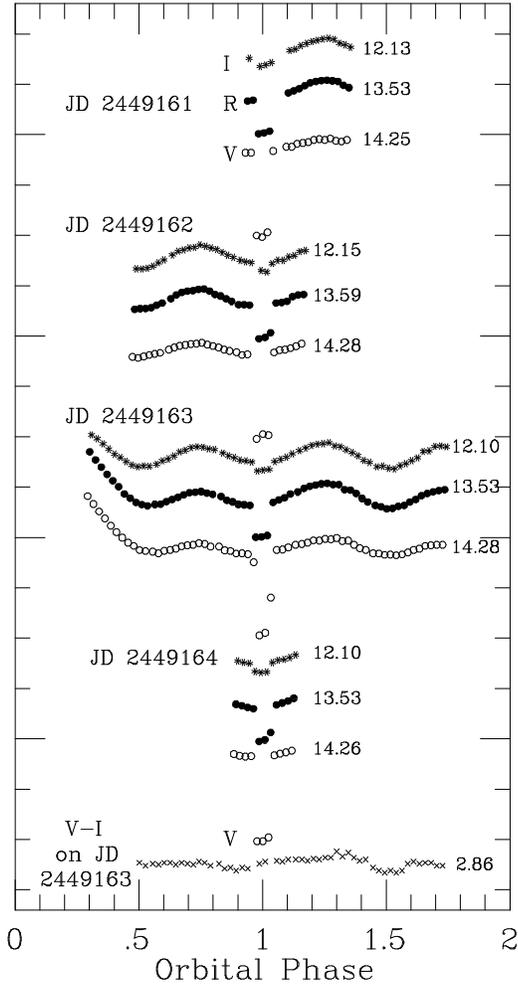,height=13cm,angle=0} 
   \end{tabular}
   \end{center}
   \caption[example]  
   { \label{fig:fig9}	   
   CCD $VRI$ photometry on HJD 2449161, 2449162, 2449163 and 2449164. 
   Asterisks indicate $I$ magnitudes, filled circles $R$ magnitudes 
   and open circles $V$ magnitudes. Orbital phase with respect to the 
   ephemeris in Table 2 is indicated on the abscissa. Ordinate carets are
   spaced at 0.5 mag intervals. The zero point for each filter subset 
   is different but can be deduced from the magnitude indicated next to the
   last point in each filter subset. The lowest curve (crosses) shows the
   $V-I$ colour on HJD 2449163. Note the decaying flare at the start of 
   the data on HJD 2449163. 
   } 
   \end{figure}  

\subsection{CCD photometry}

CCD $VRI$ photometry was obtained over the four nights 1993 June 22-25 (JD
2449161-2449164) on the 1-m telescope using the UCL camera with the RCA 
chip. Repeated cycles of $V$, $R$ and $I$ exposures, with respective
integration times of 100, 80 and 60 s were obtained. Preflash and flat field 
calibration frames were also obtained (the latter from the twilight sky in 
photometric conditions). 

Observations on the first three nights were hampered by cirrus so 
differential magnitudes for EC13471--1258 with respect to two local comparison 
stars on the frame were obtained. The fourth night was clear: E-region standard 
stars (Menzies et al. 1989) were observed and enabled transformation of the 
instrumental magnitudes of the comparison stars and EC13471--1258 to the 
$V(RI)_{\rm C}$ system. We estimate the uncertainty on the photometry 
is $\pm0.03$ mag. About 10.6 hr of useful data were obtained. Once again,
in order to assist modelling of the binary, the $VRI$ photometry are listed
in full in Table 6.

The results are plotted against orbital phase (from Table 2) in Fig. 9.
Most obvious is the previously mentioned ellipsoidal variation and the
eclipses. The eclipse depths decrease with wavelength: in $VRI$ the depths 
are respectively 53, 26 and 7 per cent. Note the decay of a flare at the 
start of the run on JD 2449163: the decline lasted $\sim1$ hr, similar to 
the flare in Fig. 5 (upper).

The ellipsoidal variation was determined by fitting a sinusoid with half
the orbital period to the data with the eclipses and the flare excised.
After correcting for the white dwarf contribution, the amplitude (peak to
peak) of the fitted sinusoid in $V$, $R$ and $I$ was respectively 27, 25 
and 22 per cent. The $V$ amplitude from the photoelectric photometry was 
22 per cent. However, there are short term variations in the ellipsoidal
variation: for example, in Fig. 9, on JD2449163, the first hump after
the flare is of lower amplitude than the second hump.

\begin{table}
\caption{$VRI$ magnitudes and colours of the component stars}
\begin{tabular}{@{}cccccc}
 $V_{\rm tot}$ & $V-R_{\rm tot}$ & $V-I_{\rm tot}$ & $R_{\rm tot}$ & $I_{\rm tot}$\\
 14.37 & 0.71 & 2.13  & 13.66 & 12.24 \\
 14.37 & 0.69 & 2.10  & 13.68 & 12.27 \\
 14.34 & 0.67 & 2.09  & 13.67 & 12.25 \\
 14.36 & 0.68 & 2.11  & 13.68 & 12.25 \\
{\bf 14.36} & {\bf 0.69} & {\bf 2.11}  & {\bf 13.67} & {\bf 12.25} \\
\\
 $V_{\rm WD}$ & $V-R_{\rm WD}$ & $V-I_{\rm WD}$ & $R_{\rm WD}$ & $I_{\rm WD}$\\
 15.04 & -0.10 & -0.37 : & 15.14 & 15.41 : \\
 15.05 & -0.06 & +0.03 : & 15.11 & 15.02 : \\
 15.03 & -0.12 & +0.03 : & 15.15 & 15.00 : \\
 15.01 & -0.12 & +0.01 : & 15.13 & 15.00 : \\
{\bf 15.03} & {\bf -0.10} & {\bf -0.08 :} & {\bf 15.13} & {\bf 15.11 :} \\
\\
 $V_{\rm Sec}$ & $V-R_{\rm Sec}$ & $V-I_{\rm Sec}$ & $R_{\rm Sec}$ & $I_{\rm Sec}$ \\
 15.21 & 1.23 &  2.91 & 13.98 & 12.30 \\
 15.20 & 1.18 &  2.84 & 14.02 & 12.36 \\
 15.16 & 1.17 &  2.81 & 13.99 & 12.34 \\
 15.22 & 1.21 &  2.88 & 14.01 & 12.34 \\
{\bf 15.20} & {\bf 1.20} & {\bf 2.86} & {\bf 14.00} & {\bf 12.34} \\
\\
 9.07 & 0.96 & 2.02 &  8.11 & 7.05 & GJ 514\\
11.43 & 1.16 & 2.71 & 10.27 & 8.72 & GJ 486\\
11.20 & 1.59 & 3.68 &  9.61 & 7.52 & GJ 551\\
\end{tabular}
\end{table}

Four eclipses were observed during the $VRI$ CCD photometry and this
allowed separation of the magnitudes and colours of the component
stars; the results are shown in Table 7 along with the means of the
four determinations. The $I$ and $V-I$ results for the white dwarf
are uncertain because it contributes less than 10 per cent in the
$I$ pass band. Note that the V magnitude of the white dwarf as
determined from the CCD data is 0.1 mag brighter than determined by
the $UBV$ data (Table 4). We felt unwilling to force agreement between
the two: the difference can be regarded as an estimate of the total
external error of the measurements. We prefer the UBV determination
as the process of multicolour photometry was much better calibrated
for the photoelectric data compared to the CCD data. 

Also shown in Table 7 are our measurements, using the same equipment,
of the colours of three nearby M dwarfs: GJ 486, 514 and 551. A full 
discussion of the secondary star will be given later but the results in 
Table 7 show that the M dwarf in EC13471--1258 is similar, at least as 
far as broad band colours are concerned, to GJ 486.

\section{Spectroscopy}

\subsection{Ultraviolet Spectroscopy}

In order to determine the effective temperature and gravity of the white
dwarf, as well as measure its radial velocity curve, {\em HST}/STIS 
observations of EC13471--1258 were obtained during a single visit lasting 
3 orbits on 28 August 1999. All observations were made through the 52" x 0.2"
aperture. In the first orbit, three spectra of 300 sec exposure time were
obtained with the G140L grating with resolution of $\sim1.5$ \AA\ and
wavelength coverage from 1150 to 1700 \AA. The remainder of the first
orbit was used to obtain two 300-sec spectra with the G230L grating
covering the wavelength range 1600--3200 \AA\ with 3.3 \AA\ resolution.
In each of the second and third orbits, six 300-sec spectra were
obtained with the G140L grating in the same manner as in the first
orbit.

   \begin{figure} 
   \begin{center}
   \begin{tabular}{c}
   \psfig{figure=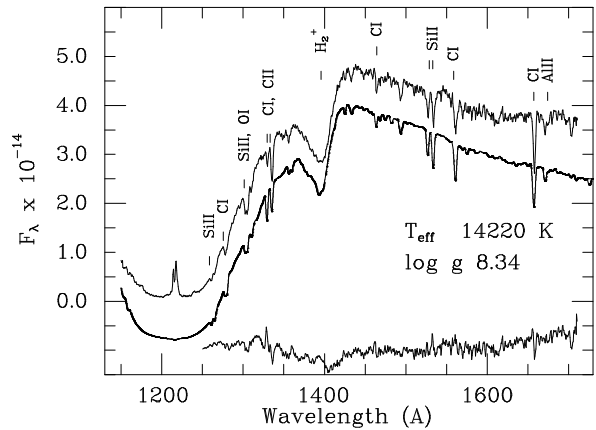,height=5.6cm,angle=0} 
   \end{tabular}
   \end{center}
   \caption[example]  
   { \label{fig:fig10}	   
   Upper curve: sum of G140L spectra of EC13471--1258 on 1999 Aug 28. 
   The individual spectra were velocity shifted to the white dwarf rest
   frame prior to summation. Various absorption features are marked. 
   Middle curve: best fit model with T$_{\rm eff}$ 14220 K and log g of 
   8.34. The model has been offset downwards by $0.8 \times 10^{-14}$ 
   with respect to the observations for clarity. Lower curve: residuals 
   of the fit in the sense of (data - model). The residuals have also 
   been shifted downwards $1.0 \times 10^{-14}$ for clarity. Further
   details are given in the text. The abscissa is in \AA\ and the ordinate 
   is in erg s$^{-1}$ cm$^{-2}$ \AA$^{-1}$.}
   \end{figure}  

With the exception of CI at 1660 \AA, and the MgII doublet at 2800 \AA, the 
G230L data are featureless. In contrast, the uppermost curve in Fig. 10 shows 
the sum of the fifteen G140L spectra. Prior to summation, the spectra were 
shifted into the rest frame of the white dwarf using the radial velocity 
solution determined below. This procedure avoided the small but detectable 
smearing of the spectrum due to the white dwarf's orbital motion. The summed
spectrum shows a continuum rising to short wavelengths but interrupted by 
very strong Ly$\alpha$ absorption (with narrow geocoronal emission in the 
core). This, along with the quasimolecular H$_2^+$ feature at 1400 \AA\ 
(Allard \& Koester 1992) indicates that the white dwarf is cool (well under 
20000 K). In addition, there are narrow metal absorption lines, mostly due to
CI and SiII. The middle curve in Fig. 10 shows our best-fitting model as
described in the next subsection; it has been displaced downwards by 
$0.8 \times 10^{-14}$ for clarity. Although a good match to the observations 
is evident, the fit is not perfect. We defer detailed discussion of the
inadequacies of the fit until the end of the next subsection.

\subsubsection{Modelling The Spectrum of the White Dwarf}

The effective temperature, surface gravity, chemical abundance and rotational
velocity of the white dwarf were determined by the classical technique of
model atmosphere calculation and fitting of the synthetic spectrum to the
observational data.

The theoretical models we employed in these calculations use the same 
procedures and programs developed over many years and applied to all 
spectral classes of white dwarfs by the Kiel group. The codes and the input 
physics are very similar to the description in Finley et al. (1997). A 
minor change compared to most applications is the chemical composition. As 
the UV spectrum shows metal lines, we have computed a new grid of 
hydrogen-rich models with metals in relative solar abundance. 

In order to limit the size of the grid of models, some initial estimates of
the atmospheric parameters were made by trial and error. As mentioned before,
the presence of the quasi-molecular H$_2^+$ feature at 1400 \AA, as well as
the absence or marginal detection of the H$_2$ feature at 1600 \AA\ enabled
us to set upper and lower bounds to the required temperature range. An initial
estimate of the line broadening was determined by convolving the theoretical
spectra with Gaussians of varying FWHM. A value of $\sim3$ \AA\ gave a good
initial fit. With this in hand, it rapidly became apparent that a solar metal 
abundance resulted in metal lines far too strong compared to the observations;
on the other hand, an abundance of 1/100 solar produced lines of insufficient 
strength. The grid that was computed, therefore, had effective temperatures in 
the range 13500 to 16000 K in steps of 250 K,  surface gravities, log g, in 
the range 7.5 to 8.5 in steps of 0.25, and metal abundances of 1/3, 1/10, 1/30, 
and 1/100 solar. 

The best model fit within this grid was determined with a Levenberg-Marquardt 
$\chi^2$ algorithm (Press et al. 1992). Our version of the method is further 
described in Homeier et al. (1998). In order to restrict the range of
possible models even further, we constrained the fit to be consistent with
the $V$ magnitude of the white dwarf, derived from both the photoelectric 
and CCD photometry during eclipse (Tables 4 and 7). We used the mean of the 
two measurement techniques $V = 15.08$ and assigned an error estimate of 
$\pm0.05$ mag. The magnitude was converted to flux using the zero point 
in Bessell (1979).

The atmospheric parameters so derived were T$_{\rm eff}$ = 14220~K and log g = 
8.34, mainly determined by the slope of the UV flux, the optical V magnitude, 
and the broad features of the Lyman $\alpha$ line and its 1400 \AA\ satellite. 
The weak metal lines do not significantly influence this determination; we 
used the grid with 1/30 solar abundances, except for Si, which was decreased 
by an additional factor of approximately 2. We discuss uncertainties in these 
estimates in the next subsection.

The modelling process also showed that the widths of the metal absorption lines 
could not be explained by the $\sim1.5$ \AA\ instrumental resolution alone.
It is, of course, possible that the instrumental resolution is somewhat worse
than nominal. However, if present, any degradation can only be modest as the 
radial velocity accuracy of the results in the next subsection would not have 
been possible if the instrumental resolution was severely degraded. The most 
obvious explanation for the additional broadening required is rotational motion 
of the white dwarf. This was investigated by first broadening the best model 
with Doppler profiles of 0, 300, 400, 500, 600, 700 and 800 km s$^{-1}$. 

Rotational broadening of spectral lines is described by a convolution
of the intrinsic profile with the broadening function 
$$
A(x) = \frac{ \frac{2}{\pi} \sqrt{1 - x^2} + \frac{\beta}{2}\,(1 -
  x^2)}{1 + \frac{2}{3}\,\beta}
$$
(see e.g. Uns\"old 1968). Here, the variable $x$ is the distance from the 
line center in units of the rotational broadening
$$
 x = \frac{\Delta\lambda}{\Delta\lambda_{rot}} \mbox{\quad with \quad} 
  \Delta\lambda_{rot} = \lambda\,\frac{V\,\sin i}{c}
$$
$\beta$ is the limb darkening coefficient with the angle-dependent intensity 
written as 
$$ I(\mu) = I(0)\,(1 + \beta \mu)$$
with $ \mu = \cos\theta$, and $\theta$ the angle between the line-of-sight
and the normal to the surface.

Our code is able to calculate the intensity for different angles from
the final model, in addition to the emerging flux.  Such a calculation
could be used to estimate $\beta$. There are two difficulties with
this direct approach: first, the program cannot calculate the
intensity directly at the limb $\mu = 0$, and second, the limb
darkening turns out to be strongly non-linear in the UV part of the
spectrum. We have therefore estimated $\beta$ by using the outermost
point calculated on the disk ($\sim 0.04$) as an approximation for the
limb value and calculating the slope from the intensity at this point and 
at the center of the disk ($\mu = 1$). The values for $\beta$ decrease from
$\sim 33$ near 1150\AA, to 7 at 1800\AA\ and 0.4 in the red at
8000\AA, with stronger variations through the profiles of stronger lines.
The classical value for Eddington limb darkening for a source function
increasing linearly with optical depth is 1.5. Fortunately, the
rotational profile (for an infinitely sharp line) does not change very
much for $\beta$ changing from $0$ to $\infty$, as can be seen from
Fig. 168 in Uns\"old (1968). This was confirmed by our own tests, 
comparing a convolution with $\beta = 1$ with one with the value 
appropriate for the observed STIS spectral range of $\beta = 15$. 

We then convolved each of these spectra with a Gaussian of FWHM equal to 
the instrumental resolution of $\sim$1.5 \AA. The resulting spectra were 
both subtracted and divided into the observational data and the resulting 
difference and quotient spectra examined (a more formal procedure is not 
justified given the difficulties in detailed matching discussed in the next 
subsection). If the model lines were too narrow, these spectra contained 
residual absorption; if too deep they showed features resembling a sinc 
function (sin x/x). Our best estimate of V$_1$~sin~i is $400 \pm 100$ km 
s$^{-1}$.

\subsubsection{Error Estimates}

Of equal importance to the atmospheric parameters of the best-fitting model
is an estimation of the uncertainties, especially for the effective temperature
and surface gravity. Accordingly, this issue was investigated in some detail. 

The residuals between the best-fitting model derived in the previous subsection
and the summed G140L data (shifted to remove the orbital motion of the white 
dwarf) are shown as the lowest curve in Fig. 10. The residuals have been 
shifted down by $1.0 \times 10^{-14}$ for clarity, and are characterized by:

\begin{itemize}
\item Low frequency trends of up to several per cent on scales of tens or 
      hundreds of \AA, most obvious at 1400 \AA\ and at the red end of the 
      spectral range shown.
\item High frequency features associated with the metal absorption lines.
\end{itemize}

We attribute the low frequency trends to a combination of (i) an imperfect fit 
of the quasimolecular H$_2^+$ feature at 1400 \AA. Despite the best available 
input physics, we were unable to achieve a better fit than shown in Fig. 10; 
and (ii) imperfections in the STIS flux calibration, especially at the red 
end of the G140L spectral range. (Note that we do not believe that the rise 
in flux at the red end is due to the quasimolecular H$_2$ feature at 1600 \AA). 

In regard to the poor fits to the metal absorption lines, the worst case 
involves the CI and II lines at $\lambda$ 1330 and 1335 \AA\ where our model 
has about equal strength for these features. In the data, however, the CI 
$\lambda$1330 \AA\ line is much weaker. On the other hand CI $\lambda$1657 
\AA\ is stronger in the data than in our model. Despite checks of {\em gf} 
values for these features, we could find no reason to change the theoretical 
{\em gf} values so these disagreements remain unexplained. The SiII lines at 
1530 \AA\ were also found to be too strong but this could be rectified by 
reducing the Si abundance by a further factor two. We suspect that this is 
a real effect but are hesitant to be adamant on this point, given the 
disagreement between theory and observation for the CI,II lines.

   \begin{figure}
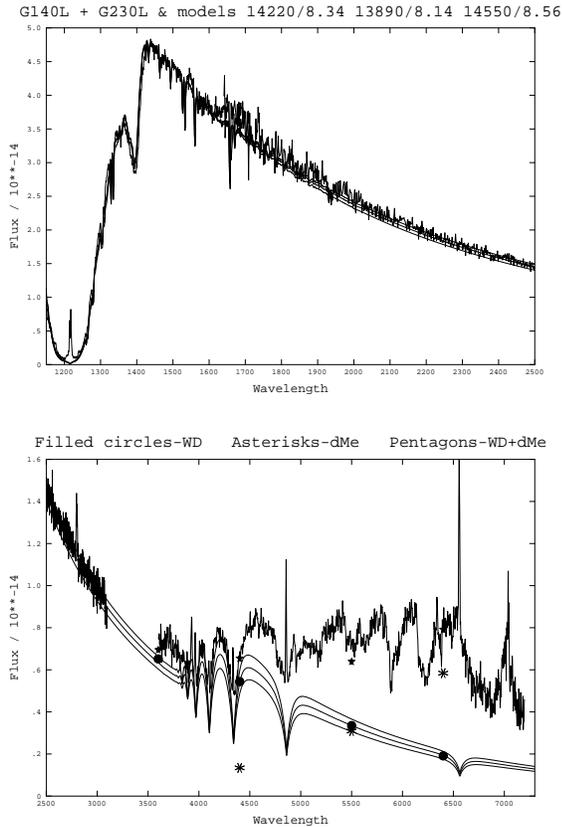
 
   \begin{tabular}{l}
   \psfig{figure=fig11a.ps,height=5.6cm,angle=270} 
   \end{tabular}
   \begin{tabular}{l}
   \psfig{figure=fig11b.ps,height=5.6cm,angle=270} 
   \end{tabular}
   \caption[example]  
   { \label{fig:fig11}	   
   Upper: sum of G140L and G230L spectra of EC13471--1258 on 1999 Aug 28, 
   along with three models with effective temperatures and surface
   gravities of (13890, 8.14), (14220, 8.34), (14550, 8.56). Lower: as for 
   the upper panel except that the wavelength range has been shifted to the
   optical. In addition, the optical spectrum from Fig. 2 is shown along
   with UBVR measurements for the white dwarf (filled circles), M dwarf 
   (asterisks) and their sum (pentagons). See text for further details. 
   The abscissa is in \AA\ and the ordinate is in erg 
   s$^{-1}$ cm$^{-2}$ \AA$^{-1}$.
   }
   \end{figure}  

The residuals are therefore not dominated by statistical errors but instead by
systematics due to inadequacies in both the calibration of the data and the 
input physics of the theoretical models. Therefore the formal errors emerging 
from the standard Levenberg-Marquardt algorithm will underestimate the errors 
of the atmospheric parameters. In addition, for white dwarfs this cool there 
is a well known correlation between effective temperature and surface gravity: 
a model with a given effective temperature and surface gravity is 
indistinguishable (at the level of the residuals in Fig. 10) in its relative 
flux distribution to another model with both a higher temperature and higher 
surface gravity, or both a lower temperature and lower surface gravity.

In order to arrive at an error estimate, we therefore employed the $V$ 
magnitude of the white dwarf as a discriminator. The upper panel of Fig. 11 
shows a comparison between the sum of the G140L and G230L data and three 
models, one of which is the best fitting model (14220, 8.34) shown in Fig. 
10. The other two models have atmospheric parameters of 13890 and 14550 K, 
with corresponding surface gravities of 8.14 and 8.56, respectively. All 
three models have been normalized to the G140L observations at 1450 \AA\ and
all three provide a reasonably close match over the G140L wavelength range 
(due to the correlation between temperature and surface gravity mentioned 
above). Note, however, that the G230L data lie above all three models in 
the range 1600--2150 \AA. We attribute this discord to STIS calibration 
uncertainties.

The lower panel of Fig. 11 shows the same three models as in the upper
panel but extended to the optical region. The sum of the G230L data are
shown, along with the optical flux distribution from Fig. 2, and the
$UBVRI$ photometry of the white dwarf (filled circles), M dwarf (asterisks)
and the combination (pentagons) (Tables 4 and 7). The photometry was
converted to fluxes using the zero points in Bessell (1979). The 
best-fitting model, (14220, 8.34), passes through the $V$ measurement at 
5500 \AA\ with the (13890, 8.14) model lying above and the (14550, 8.56) 
model lying below. We emphasize that none of the data in Fig. 11 has been
adjusted for consistency but simply taken from the measurements and
associated calibrations. The ability of the photometry to distinguish
the different models relies on the absolute calibration of the G140L
data and the $V$ measurement. It is our belief that the relatively
poorer fit to the $V$ measurement of the (13890, 8.14) and (14550, 8.56)
models represents a reasonable estimation of $\pm1 \sigma$ for the atmospheric 
parameters. For comparison with the $\chi^2$ technique, the models bracketing 
the best fit in Fig. 11 fall on the $\Delta \chi^2 = 2.3$ contour of the error 
ellipse (Fig. 15.6.4 of Press et al. 1992). We stress the correlation of the 
errors so that if the real effective temperature of the white dwarf is either 
higher or lower than 14220 K, the surface gravity must correspondingly be 
adjusted up or down.

As mentioned already, the metal lines play no role in the effective
temperature and surface gravity determination and as the fits were
imperfect, it was deemed acceptable to judge the fitting by eye. 
A reasonable overall fit was obtained with abundances of 1/30 solar 
(except that Si was decreased by a further factor two); 1/10 as well 
as 1/100 solar models were clearly less satisfactory. We deduce from 
this a conservative error estimate of $\pm0.5$ dex.

In summary, the modelling of the white dwarf spectrum has yielded
the following results:

\begin{tabular}{rcl}
T$_{\rm eff}$ & = & $14220 \pm 350$ K\\
  log g       & = & $8.34 \pm 0.22$ \\
  log Z       & = & $-1.5 \pm 0.5$ 
\end{tabular}

\begin{table}
\caption{White Dwarf Radial Velocities}
\begin{center}
\begin{tabular}{@{}cccc}  
Orbital &   Heliocentric   & Orbital &  Heliocentric    \\
Phase   & Radial Velocity  & Phase   & Radial Velocity  \\
        &  (km s$^{-1}$)   &         &  (km s$^{-1}$)   \\
\\
  0.767   &  176   &  1.286   &  -26 \\
  0.801   &  214   &  1.599   &  131 \\
  0.826   &  253   &  1.624   &  115 \\
  1.154   &  -71   &  1.666   &  129 \\
  1.187   &  -80   &  1.690   &  165 \\
  1.212   &  -78   &  1.706   &  207 \\
  1.237   &  -76   &  1.740   &  234 \\
  1.261   &  -80  \\
\\
\\
Template & (K$_1$\ sin\ i)   & $\gamma_{1,\odot}$ \\
         & (km s$^{-1}$)     & (km s$^{-1}$) \\
Unshifted Sum &   96 $\pm$  8\\
Shifted Sum   &  124 $\pm$  9\\
Model         &  137 $\pm$ 10  &  61 $\pm$ 10 \\
\end{tabular}
\end{center}
\end{table}

   \begin{figure} 
   \begin{center}
   \begin{tabular}{c}
   \psfig{figure=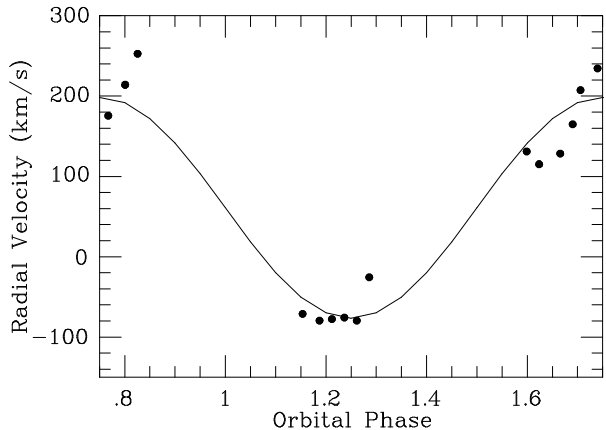,height=5.6cm,angle=0} 
   \end{tabular}
   \end{center}
   \caption[example]  
   { \label{fig:fig12}	   
   White dwarf radial velocities from the HST spectra relative to the model 
   template (filled circles) with best fitting orbital solution superimposed 
   (solid line).}
   \end{figure}  

\subsubsection{White Dwarf Radial Velocities}

One of the prime motivations for obtaining HST ultraviolet spectroscopy
was to measure the radial velocity variations of the white dwarf arising
from the binary motion. The fifteen G140L spectra were timed to occur
as close to binary quadrature as possible, given the limited time available
and the constraints of the HST orbit. Radial velocities were extracted by
cross-correlating the fifteen individual spectra with a template (Tonry
\& Davis 1979) from which the continuum and other low order variations
has been removed. Three separate templates were tried: (i) the sum of
all the G140L spectra; (ii) the best fitting model (middle curve in Fig. 
10); (iii) the sum of all the G140L spectra but with each having been 
shifted to the rest frame of the white dwarf by the preliminary solution 
to the radial velocity curve (upper curve in Fig. 10). This latter technique 
is intended to sharpen the features in the summed spectrum which would
otherwise be slightly degraded by the orbital motion. For all templates and 
individual spectra, the continuum was removed prior to cross-correlation. 
In addition, Ly$\alpha$ is too broad to contain sensible velocity information 
so this line was treated as continuum and removed as well. The resulting 
spectrum's strongest feature was the quasimolecular H$_2^+$ feature at 1400 
\AA\ with additional velocity information coming from the metal absorption 
lines.  

The resulting velocities, for the model template only, are listed in Table 8
and plotted in Fig. 12. Both the motion of the earth with respect to the 
sun and the spacecraft with respect to the earth have been accounted for 
in the values listed in Table 8 (the latter is, of course, very small: less
than 5 km s$^{-1}$). 

The orbital ephemeris was used to calculate the orbital phase of the 
velocities and a sinusoidal curve with fixed orbital period and orbital 
phase of the form:
$$
{\rm V_r\ =\ \gamma_1\ -\ (K_1\ sin\ i)\ sin( 2\pi\ \phi_{orb} ) }
$$
was fitted to the velocities by least squares (with only 15 data 
points, the best results are obtained if the fewest number of variables 
are allowed). The best fitting curve for the model template velocities 
is plotted as the solid line in Fig. 12. Table 8 also shows the orbital 
semi-amplitudes, K$_1$, for the three different templates, along with the 
formal errors of the fit. As expected, the unshifted sum template has a 
lower velocity than the other two which are consistent with each other 
within the errors of measurement. The model template velocities also 
provide a mean, or gamma, velocity for the fit, which we shall use in 
detecting the gravitational redshift of the white dwarf.

\begin{table}
\caption{Optical Spectroscopy Observing Log (superscripts in column 1
         refer to Table 10)}  
\begin{center}
\begin{tabular}{@{}cccccc}
     Date     &    Time     &    Exp.   &   Wave-    &  Resol- &  No of  \\
              &    Span     &    Time   &  length    &  ution  &  Spect. \\
              &             &           &   Range    &        \\
              &    (UT)     &   (sec)   &  (\AA)     & (\AA) \\
\\
1992 Apr\\
    02/03     & 21:14-21:57 &  2400     & 5000-7400  &  4.4   & 2  \\
              & 22:12-03:32 &   600     & 3800-5100  &  2.2   & 24 \\
    05/06     & 02:41-03:36 &  2400     & 5000-7400  &  4.4   & 1  \\
\\
1992 May\\
    09/10     & 18:02-01:34 &   600     & 5500-6700  &  2.2   & 34 \\
    11/12     & 17:40-22:35 &   900     & 3500-5200  &  3.0   & 18 \\
\\
1993 Feb\\
    24/25     & 22:55-02:55 &   900     & 3800-5100  &  2.2   & 16 \\
    25/26     & 22:59-02:46 &   900     & 3800-5100  &  2.2   & 15 \\
    26/27     & 22:51-02:50 &   900     & 3800-5100  &  2.2   & 14 \\
\\
1993 Apr\\
  14/15$^{\ 1}$ & 19:14-02:02 &   600     & 5600-6800  &  2.2   & 34 \\
  15/16$^{\ 2}$ & 21:26-21:53 &  2400     & 5000-7400  &  4.4   & 1  \\
  15/16         & 23:34-23:55 &  1200     & 3700-7100  &  6.0   & 1  \\
  16/17$^{\ 3}$ & 18:58-01:02 &   600     & 5600-6800  &  2.2   & 19 \\
  18/19$^{\ 4}$ & 19:00-02:27 &   600     & 5600-6800  &  2.2   & 36 \\
  19/20$^{\ 5}$ & 19:14-00:24 &   600     & 5600-6800  &  2.2   & 26 \\
\\
1993 June\\ 
    24        & 10:27-14:41 &   900     & 5600-6800  &  2.8   & 14 \\
    25        & 08:17-13:30 &   900     & 5000-7400  &  2.8   & 21 \\
    26        & 08:33-13:31 &   900     & 3700-7100  &  2.8   & 18 \\
    27        & 09:06-14:18 &   900     & 5600-6800  &  2.8   & 20 \\
\end{tabular}
\end{center}
\end{table}

\begin{table}
\caption{Observed M Dwarfs (superscripts in column 1 refer to Table 9)}
\begin{center}
\begin{tabular}{@{}lcccccc}
    Name             &   Spec.  &  Rad.    & V-R & V-I    & TIO5 & CAH1 \\
\ \ GJ               &   Type   &  Vel.    \\
                     &          & (km/s)   \\
\\
\ \ 514 $^{\ 2}$     &   M0.5   &          & 0.98 & 2.04  \\  
\ \ 382 $^{\ 5}$     &  M1.5-2  &          & 1.00 & 2.17  \\  
\ \ 393 $^{\ 2,5}$   &    M2    &          & 1.02 & 2.24  \\  
\ \ 381 $^{\ 2,3,4}$ &   M2.5   &  31,33   & 1.04 & 2.32  \\  
\ \ 273 $^{\ 1,2}$   &   M3.5   &   18     & 1.17 & 2.71  & 0.43 & 0.86 \\  
\ \ 486 $^{\ 2}$     &   M3.5   &          & 1.17 & 2.69  & 0.39 & 0.88 \\
\ \ 447 $^{\ 2,5}$   &    M4    &  -31     & 1.29 & 2.97  & 0.32 & 0.83 \\  
\ \ 285 $^{\ 5}$     &   M4.5   &   26     & 1.27 & 2.95  \\  
\ \ 299 $^{\ 2}$     &   M4.5   &          & 1.25 & 2.92  \\  
\ \ 551 $^{\ 1,2,4}$ &   M5.5   &          & 1.65 & 3.65  \\  
\\
M dwarf              &   M3.5-  &          & 1.20 & 2.75- & 0.32 & 0.89 \\
in 13471--           &    M4    &          &      & 2.95  \\
1258\\
\end{tabular}
\end{center}
\end{table}

\subsection{Optical Spectroscopy}

Table 9 shows an observing log for the optical spectroscopy. With the
exception of June 1993, all the rest of the data were obtained with the
Image Tube Spectrograph on the SAAO 1.9-m telescope, equipped with an 
S20 photocathode and Reticon diode array detector. Much
of the data was obtained for radial velocity measurements and great
care was taken to ensure the stability of the wavelength scale of the
spectra by observing Cu-Ar or He-Ne arcs every 20 min. Pixel to pixel
sensitivity variations in the detector were calibrated using observations 
of an incandescent lamp. Spectrophotometric standards 
were also observed to achieve flux calibration but as these were mostly 
obtained through a narrow slit, the absolute calibration is not reliable. 
One of the standards from Hamuy et al. (1994), EG99, is only $\sim7$ degrees 
away from EC13471--1258. Despite this, it is also possible that 
wavelength-dependent slit losses are present, especially at the extreme ends 
of the wavelength range covered, due to the fact that the spectrograph slit 
was aligned in the E-W direction and the image tube chain suffers some
variability in sensitivity at the edges of the detector. The data were 
reduced by flat-fielding, fitting a 5th order polynomial to the arc lines
to define the dispersion relation, subtracting the sky observed in the
second Reticon diode array 30 arcsec from the object on the sky, and flux 
calibrating with a relation derived from the standard star observations. 

Spectra were also acquired on the nights of 1993 June 24-27 at Mount 
Stromlo \& Siding Spring Observatories with the 2.3-m at Siding Spring,
equipped with the Double Beam Spectrograph. Red and blue spectra were
acquired simultaneously with the vast majority of exposures being of
900 s duration but with occasional shorter exposures of 600 and even 450
s in good conditions. PCA detectors were used in each arm of the
spectrograph. The red spectra covered the wavelength range 6420--6680 \AA\ 
with a resolution of $\sim2.8$ \AA; the blue spectra covered the wavelength
range 3900-4500 \AA\ with a resolution of $\sim3.0$ \AA. 

Only one arc per night was observed. The stability of the wavelength scale
was achieved by correcting the observed sky spectra so that the faint night
sky emission lines were placed at their correct wavelengths. In the case
of the blue spectra, the first night showed no sky lines useful for this
purpose so these data were not used where wavelength accuracy was required.

Whereas contributions to the spectrum in the HST data originate only in
the white dwarf, in the optical spectrum both white dwarf and M dwarf
contribute, in roughly equal amounts at 5500 \AA. Accordingly it is
necessary to disentangle these contributions. The relatively reliable
model for the white dwarf derived above shows a smooth flux distribution 
in the optical (Fig. 11); longward of 5500 \AA, its contribution declines
and H$\alpha$ is its only spectral feature so we decided to concentrate
on the red spectra first to measure the properties of the M dwarf. Our
approach is as empirical as possible: despite substantial progress over
the last decade in modelling all aspects of M dwarfs, there is still
discord between theory and observation (summarized in Reid \& Hawley 
2000). Accordingly, in addition to the spectrophotometric standard stars, 
M dwarfs were also observed (Table 10), for the purposes of
spectral type and radial velocity determination. The superscripts in Table
10 next to the M dwarf star names indicate that that star was observed on
the date in Table 9 with the same superscript. Spectral types were taken 
from Kirkpatrick, Henry \& McCarthy (1991), Reid, Hawley \& Gizis (1995) 
and Hawley, Gizis \& Reid (1996); for stars with multiple spectral type 
estimates, the different estimates were identical except in the case of 
GJ 382 as indicated in Table 10. Radial velocities with an accuracy of 
better than 1 km s$^{-1}$ were taken from Nidever et al. (2002) except for GJ 
381 which is a known binary (Delfosse et al. 1999); Dr. Xavier Delfosse 
kindly supplied the radial velocity appropriate for the system's very much 
brighter primary star (in the R band) at the time of our highest resolution 
observations: 31 km s$^{-1}$ on 1992 May 09/10 and 33 during April 1993. 

We also include in Table 10 VRI colours and TiO and CaH band indices which 
we shall use later. The VRI data were taken from Koen et al. (2002) or 
Bessell (1990); overall agreement between these authors is discussed in 
Koen et al. and the specific stars observed in common were different by at 
most 0.02 mag. Photometry for one object, GJ 299, was taken from Table A1 
of Reid \& Hawley (2000). The molecular band indices are defined in Table 2
of Reid, Hawley \& Gizis (1995); the data in Table 10 are our measurements 
which were made on the night of 1993 Apr 15/16; we estimate an error of
0.04 in the indices. Comparison with the corresponding measurements in
Reid, Hawley \& Gizis (1995) measurements show that our estimate of CAH1 
is systematically larger by 0.05.

\subsubsection{Radial Velocities of the M Dwarf}

The radial velocity semi-amplitude of the M dwarf is likely to be in
excess of 200 km s$^{-1}$, and its absorption lines are also likely to be
considerably broadened by rotation, so before examining its spectrum 
it is necessary to measure and account for the blurring effect of these 
motions. For this purpose, we selected the highest resolution data 
among the SAAO red spectra, obtained on 1992 May 09/10, and 1993 Apr 
14/15, 16/17, 18/19, 19/20.

Examination of these data showed that the radial velocity curve of the
M dwarf could be traced by the H$\alpha$ emission line, by the Na D 
absorption lines or by the TiO molecular bands and other weaker metal 
lines. In the case of H$\alpha$, it is possible that it will not
reflect the true motion of the centre of mass of the M dwarf because
the majority of the emission might arise on the side facing the white 
dwarf due to the heating effect of the white dwarf. We attempted to use 
the Na D lines but achieved results of marginal reliability due to the 
weakness of the lines. Instead we used the TiO bands and the forest of 
weak metal lines as the most reliable indicator of the motion of the
centre of mass of the M dwarf.

   \begin{figure} 
   \begin{center}
   \begin{tabular}{c}
   \psfig{figure=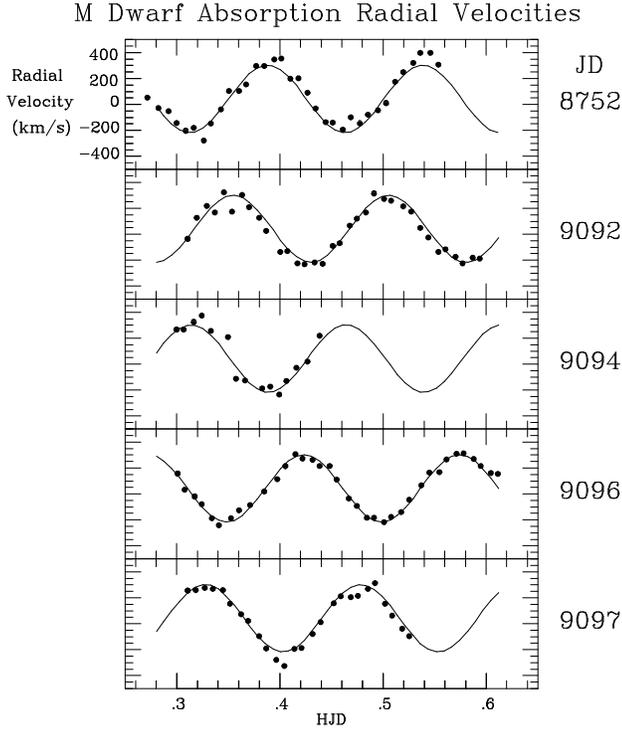,height=9.6cm,angle=0} 
   \end{tabular}
   \end{center}
   \caption[example]  
   { \label{fig:fig13}	   
   M dwarf absorption feature radial velocities relative to the template 
   GJ447 (points) on the nights of 1992 May 09/10 and 1993 Apr 14/15, 
   16/17, 18/19, 19/20 (JD 2448752, 2449092, 2449094, 2449096 and 2449097).
   The best fitting orbital solution from Table 11 is superimposed (solid 
   line).}
   \end{figure}  

Anticipating the results on the rotational velocity and spectral type
presented below, cross-correlation templates (Tonry \& Davis 1979) were 
constructed over the wavelength range 5600-6500 \AA\ using the spectra of 
the M dwarfs GJ273, GJ381, GJ285 and GJ447 which were observed with the 
same instrumental setup (Tables 9 and 10). These spectra were broadened 
by convolution with a Doppler profile with V$_2$ sin i = 125 km s$^{-1}$. The 
overall shape of the continuum was removed by subtracting a 4th order 
polynomial: this procedure retained the sharp edges of the molecular 
bands, as well as the absorption lines. 

\begin{table}
\caption{M dwarf absorption radial velocities}
\begin{center}
\begin{tabular}{ccc}
Template star & (K$_2$\ sin\ i) & $\gamma_{2,\odot}$ \\
\\
GJ381 & $272 \pm 6$ & $  0  \pm 4$ \\
GJ273 & $272 \pm 6$ & $ -23 \pm 4$ \\
GJ447 & $260 \pm 6$ & $  13 \pm 4$ \\
GJ285 & $260 \pm 6$ & $  1  \pm 4$ \\
\\
Mean  & $266 \pm 6$ & $ -2  \pm 4$ \\
\end{tabular}
\end{center}
\end{table}

The radial velocities extracted from the five nights of data were corrected 
for the earth's motion along the line of sight to the program star as well 
as the template star (Table 10). Orbital phases were calculated from the 
ephemeris in Table 2 and a function of the form:
$$
{\rm V_r\ =\ \gamma_2\ +\ (K_2\ sin\ i)\ sin( 2\pi\ \phi_{orb} ) }
$$
was fitted by least squares. The results for the four template stars are 
listed in Table 11 and, in the case of GJ447, plotted in Fig. 13 as a
typical illustration of the quality of the results. There is a tendency
in Table 11 for the earlier template stars to give a larger radial
velocity amplitude. As shown below, it is difficult to pin down the
spectral type of the M dwarf in EC13471--1258 to better than 0.5 spectral
subclasses, if for no other reason than the spectral type varies judging
from the $V-I$ colour variation. We believe that the uncertainties are 
dominated by the choice of template star; we therefore adopt a conservative
estimate for the uncertainties and adopt the same error estimate for the
estimate of the semi-amplitude and mean or $\gamma$ velocity as for the
individual measurements. The results are $266 \pm 6$ km s$^{-1}$ and $-2 
\pm 4$ km s$^{-1}$, respectively.

   \begin{figure}
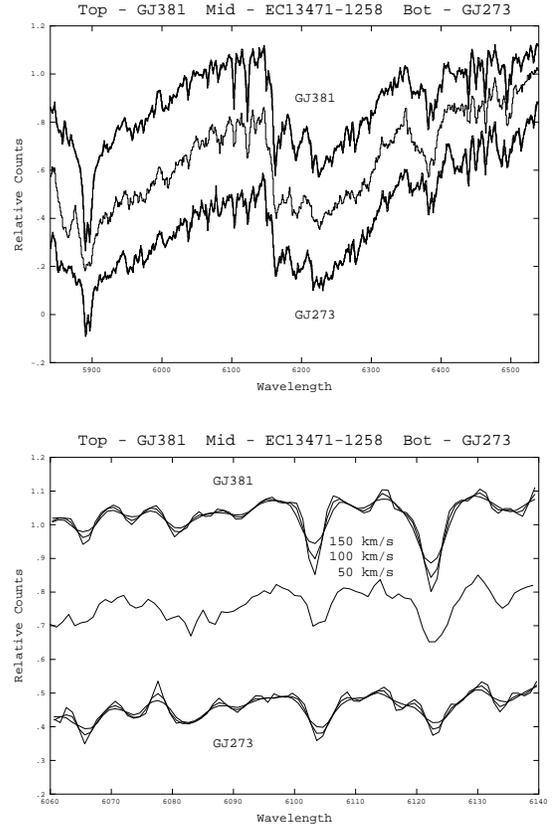
 
   \begin{tabular}{l}
   \psfig{figure=fig14a.ps,height=5.6cm,angle=270} 
   \end{tabular}
   \begin{tabular}{l}
   \psfig{figure=fig14b.ps,height=5.6cm,angle=270} 
   \end{tabular}
   \caption[example]  
   { \label{fig:fig14}	   
   Upper: The top curve is the sum of the 2.2 \AA\ resolution red spectra 
   from 1993 Apr 16/17 and 18/19 of GJ381. The bottom curve is the 2.2
   \AA\ resolution red spectrum of GJ273 obtained on 1993 Apr 14/15.
   The middle curve is the sum of the 2.2 \AA\ resolution red spectra of
   EC13471--1258, shifted into to the red frame of the M dwarf before
   summation, and with the contribution of the white dwarf subtracted.
   All spectra have been left in extinction corrected counts (to avoid 
   amplifying the noise at the ends of the spectral range) and normalized 
   to 1.0 just shortward of H$\alpha$. The spectra of GJ381 and GJ273 have 
   been shifted vertically upwards and downwards for clarity. Lower: as in 
   the upper curve but with the wavelength scale expanded around 6100 \AA.
   In addition, the original spectra of GJ381 and GJ273 have been
   convolved with Doppler profiles with V sin i = 50, 100 and 150 km s$^{-1}$
   to enable the rotational velocity of the M dwarf in EC13471--1258 to
   be estimated.
   }
   \end{figure}  

\subsubsection{Spectral type and effective temperature of the M dwarf}

There are a variety of approaches for determining the spectral type of
the M dwarf in EC13471--1258:

\begin{itemize}
\item The least squares fitting approach of Kirkpatrick, Henry \&
      McCarthy (1991). We used this technique on the 4.4 \AA\ resolution
      red spectra from 1993 Apr 15/16 and determined that the spectral
      type was later than that of GJ381 (M2.5) and earlier than GJ299
      (M4.5). It was difficult to distinguish among the spectral
      types in between. Similar conclusions can be drawn from the higher
      resolution (2.2 \AA) red spectroscopy from 1993 Apr: Fig 14a shows
      a comparison of the spectra of GJ381 (M2.5)(top), GJ273 (M3.5)(bottom)
      with that of the M dwarf in EC13471--1258 (middle). The radial
      velocity smearing and contribution of the white dwarf has been 
      removed from the EC13471--1258 spectrum while the spectra of the 
      comparison stars have been broadened with a Doppler profile of 125
      km s$^{-1}$ (see discussion below on this point). It is apparent that 
      the Na D absorption is not nearly as strong in EC13471--1258 as in
      GJ381. The molecular bands in the latter have smaller equivalent
      width too.
\item The narrow band spectrophotometry technique of Reid, Hawley \&
      Gizis (1995): see their fig.2 and table 2. The TIO5 index measures 
      the band head depth of the TIO band at 7042 \AA\ and is the best 
      estimate of spectral type. In the case of EC13471--1258, this 
      indicates a spectral type of M4.5. However, the index is very 
      sensitive to the contribution of the white dwarf. This in turn 
      depends on the temperature of the white dwarf: as the contribution
      of the white dwarf to be subtracted decreases, the index gets larger
      and the spectral type earlier. We estimate that the measured TIO5 
      has an uncertainty of about 0.05 due to this effect and so this index 
      does not provide greater discrimination than the Kirkpatrick et 
      al. (1991) technique.
\item Probably the most sensitive indicator of spectral type and temperature
      in mid M dwarfs is $V-I$ colour. Indeed, a plot of this colour against
      spectral type for the catalog data presented in Reid et al. (1995)
      and Hawley et al. (1996) shows a tight linear correlation about
      the point ($V-I$, SpType) = (2.7, 3.5) with a slope of 2.5 (i.e. stars
      with spectral type M3 (or M4) have $V-I$ of 2.5 (or 2.9) respectively.
      Inspection of Table 10 confirms this impression. $V-I$ colours for
      the M dwarf (i.e. with the white dwarf subtracted) were calculated
      from the CCD photometry (Table 6) yielding a mean value of 2.86
      and a range from 2.95 to 2.75. A plot of the data on HJD 2449163
      is shown as the lowest curve (crosses) in Fig. 9. From this we
      conclude that the spectral type of the M dwarf in EC13471--1258
      varies between M3.5 and M4. This conclusion is consistent with the
      previous two bullets and all other information at our disposal.
      It is interesting to note that the variation in $V-I$ in Fig. 9
      is much smaller than the modulation in $V$. 
\end{itemize}

   \begin{figure*}
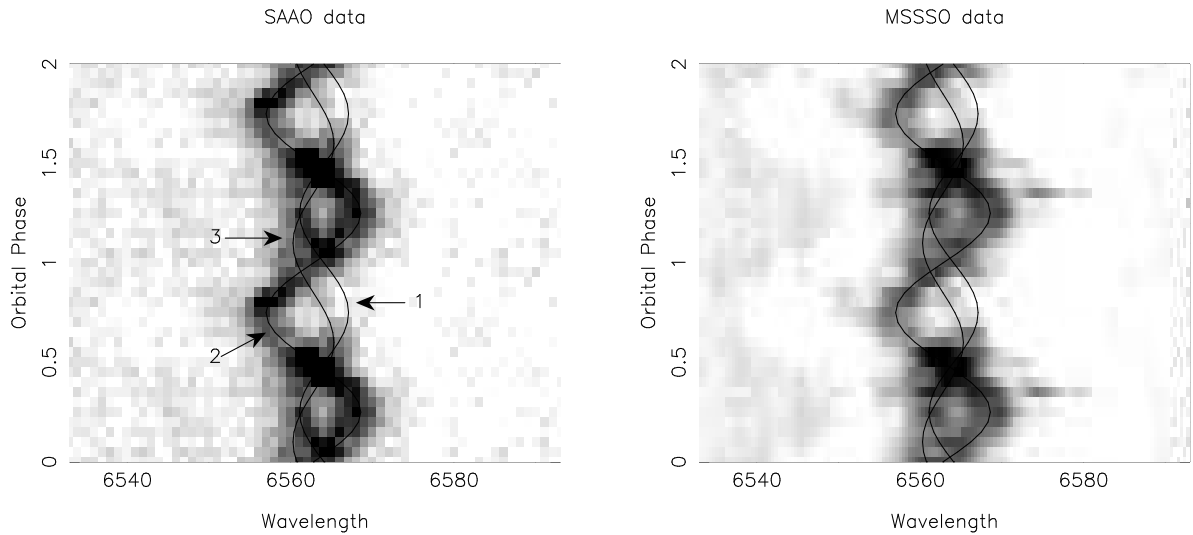
 
   \begin{tabular}{l}
   \psfig{figure=fig15a.ps,height=7cm,angle=0} 
   \end{tabular}
   \begin{tabular}{l}
   \psfig{figure=fig15b.ps,height=7cm,angle=0} 
   \end{tabular}
   \caption[example]  
   { \label{fig:fig15}	   
   Left: Trailed SAAO spectra of the H$\alpha$ emission line, folded on 
   the orbital ephemeris. Two cycles are shown for the sake of continuity,
   the second cycle being a copy of the first. Each spectrum was scaled
   so that the continuum just outside the H$\alpha$ profile is normalized
   to unity. The grey scale is chosen so that white corresponds to a value 
   of 1 (i.e. the continuum) and black to 2.2 (the peak of the emission
   profiles in all spectra was 2.5). Orbital phase is shown on the 
   ordinate axis and wavelength is shown on the abscissa. The curve
   labelled 2 corresponds to the expected motion of the M dwarf as
   deduced from the absorption line radial velocity curve derived above.
   The curve labelled 1 corresponds to the expected motion of the white 
   dwarf as derived from the velocity variations in the HST spectra; the
   gamma velocity of 60 km s$^{-1}$ is {\em included} in curve 1. The curve 
   labelled 3 corresponds to the motion of the feature in the Doppler
   tomogram at velocity co-ordinates (75, -75) km s$^{-1}$. It can be seem 
   as a faint feature visible between orbital phase 0.6 and 0.9 and lying 
   to the blue of the expected motion of the white dwarf. See text for 
   more details. Right: As in the left panel except that the MSSSO data 
   are shown. 
   }
   \end{figure*}  

From table 4.1 or fig. 4.13 of Reid \& Hawley (2000), we derive a 
mean temperature for the M dwarf of $3100 \pm 50$ K with intrinsic 
variation of semi-amplitude 75 K about this mean. Whether or not
this variation is simply a function of orbital phase, or whether it
is due, for example, to slowly migrating spotted regions on the surface
of the M dwarf, will require far more data than is available in this
study.

\subsubsection{Rotational velocity of the M dwarf}

Fig. 14b shows the same spectra as in Fig. 14a but with an expanded
wavelength scale and a number of different rotational models. It is
clear that both rotation and spectral type contribute to the depths
of the lines. If GJ381 is used as the template, a rotational velocity
of 150 km s$^{-1}$ is deduced. However, as shown in the last section, GJ381
is too early. The spectral fit to GJ273 (lower spectra) in parts of
the wavelength range covered is too poor to judge the rotational
velocity. No objective technique could be devised and inspection was
resorted to. We conclude that the rotational velocity of the M dwarf
is ${\rm V_2\ sin\ i\ =\ 125\pm25\ km s^{-1}}$.

\subsubsection{Metallicity of the M dwarf}

With considerably subsolar metal abundances in the white dwarf, it is of
some interest to determine the composition of the M dwarf. As illustrated
in fig. 2.18 of Reid \& Hawley (2000), chemical composition in M dwarfs
is indicated by the increase in strength of the CaH compared to the TiO 
molecular bands as metal abundance decreases. Quantitative measurement 
is described by Gizis (1997); as shown by his fig. 1a, the distinction 
between disk M dwarfs on the one hand and subdwarfs or extreme subdwarfs 
on the other (with metal abundance of 1/16 solar or less) is clearest in 
the CAH1-TIO5 diagram. The measurements of these band indices rely on 
narrow band spectrophotometry and in order to reduce systematic
uncertainties in our spectra of EC13471--1258, we measured these band
strengths for three M dwarfs of similar spectral type (Table 10). As 
discussed above, for the M dwarf in EC13471--1258, a reasonably accurate 
subtraction of the spectrum of the white dwarf was required and additional 
uncertainty is associated with this. 

The indices for the M dwarf in EC13471--1258 match those of the similar
stars (all of which are normal disk M dwarfs); moreover, even after 
correcting for the systematic overestimation of the CAH1 index, it 
clear that the indices for the M dwarf in EC13471--1258 lie well away 
from the metal deficient M dwarf region in fig. 1a of Gizis
(1997). We conclude that the M dwarf in EC13471--1258 has normal solar
abundance and defer until later discussion of the considerably lower metal 
abundance in the white dwarf.

\subsubsection{The behaviour of the emission lines}

It was obvious even at the telescope that the profiles of the Balmer 
emission lines were comprised of more than one component; this constrained us 
from using them for radial velocity measurements as it would not be apparent 
how to interpret the results. We therefore folded all the SAAO highest 
resolution red spectra into 20 phase bins using the orbital
ephemeris and scaled each in such a manner that the continuum just
outside the H$\alpha$ emission line profile was normalized to unity.
The results are shown as a trailed spectrum in the left panel of Fig. 15. 
The right panel shows the same procedure applied to the MSSSO data. 
Bearing in mind the worse resolution of the latter data, both panels are
consistent with each other, indicating that the features seen are not
transient but repeatable on a time scale of months.

The left panel shows that the H$\alpha$ emission line is indeed comprised of 
two components, leading to a doubling of the line profile at orbital phase 
0.25, and a single peak at the phases of conjunction, 0.00 and 0.50.  The 
stronger of the two components is well matched to the curve labelled 2; 
this curve is not a fit to this component but is instead the motion of the 
M dwarf derived from the absorption line radial velocity curve derived above. 

Turning to the weaker of the two components, it is apparent that this varies
around the orbital cycle, peaking in strength at orbital phase 0.20-0.25.
At orbital phase 0.75, although much weaker in strength than half a cycle
earlier/later, it is certainly present: examination of the individual
spectra showed a clear doubling of the emission line in the orbital
phase interval 0.60 to 0.70, disappearance of the weaker component
around orbital phase 0.75, a brief reappearance at orbital phase 0.80
before merging with the redward moving stronger component at eclipse.

The curve labelled 1 corresponds to the expected motion of the white dwarf 
as derived from the velocity variations in the HST spectra, including its 
gamma velocity of 60 km s$^{-1}$. We are not convinced that this curve is a 
satisfactory description of the velocity variation of the weaker component
in the data because it co-incides with no emission between orbital phase 
0.70 and 0.85. Omitting the gamma velocity (which we attribute to a 
gravitational redshift from emission near the surface of the white dwarf), 
makes the fit worse. Whatever the case, the weaker component must originate 
on the same side of the binary centre of mass as the white dwarf as its 
curvature is clearly of opposite sign to that of the M dwarf's motion.

We {\em are} convinced of the reality of the feature visible between 
orbital phases 0.60 and 0.90 just slightly to the blue of the curve labelled 
1.  The curve labelled 3 is derived from the Doppler tomogram discussed  
below. Note that its motion does {\em not} co-incide with the arc which 
is strongest between orbital phases 0.00 and 0.25 and which the curve 
labelled 1 passes through. Thus neither curves 1 nor 3 provide, on their
own, a satisfactory representation of the weaker component. 

It is also important to note the variations in strength of the stronger
component. At phases close to conjunction, there is an obvious increase
in strength which is plausibly explained by the summation of the two
components. However, there is additional intrinsic variation in the
strength of the stronger component: it weakens by at least a factor
of two in the orbital phase interval 0.60-0.70 compared to the combined 
strength at orbital phase 0.50, and strengthens again suddenly at orbital 
phase 0.75. It is clear that the profile cannot be explained by the simple
summation of two optically thin emission line sources of variable radial
velocity. Turning briefly to the blue optical spectra (Fig. 17) which will 
be discussed more fully below, the relative fluxes in the Balmer emission 
lines shows that they are optically thick: the Balmer decrement is flat - 
the line strength ratios H$\alpha$/H$\beta$, H$\gamma$/H$\beta$, and
H$\delta$/H$\beta$ are 1.4, 0.67 and 0.83, whereas Case B recombination 
predicts values of 2.80, 0.47 and 0.29, respectively (table 4 of Ferland et 
al. 1982). This provides a natural explanation for the intrinsic line strength 
variations in the H$\alpha$ profile: absorption of Balmer line photons by 
intervening gas is easily capable of explaining the variations seen. Specific 
scenarios in which this is taking place will be discussed later.

   \begin{figure} 
   \begin{tabular}{c}
   \psfig{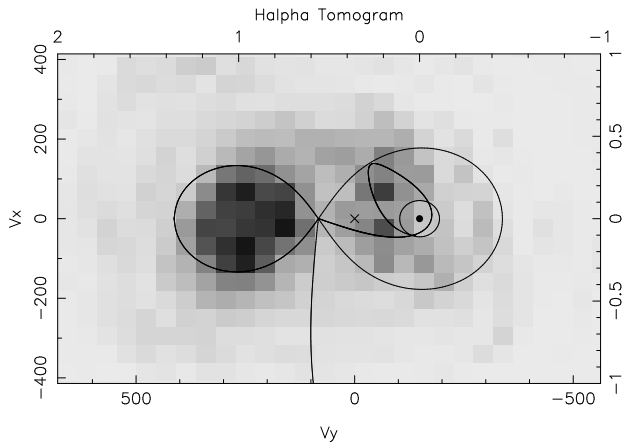} 
   \end{tabular}
   \caption[example]  
   { \label{fig:fig16}	   
   Left: Doppler tomogram of the H$\alpha$ emission line. Two scales are 
   shown: velocity (lower horizontal and left vertical) and spatial 
   (upper horizontal and right vertical). Both Roche lobes are shown
   as well as the binary centre of mass (cross), the white dwarf position
   (small filled circle) and the mass transfer stream in spatial
   co-ordinates as it loops around the white dwarf. The small circle
   surrounding the white dwarf is the inviscid disc radius (eq. 2.19 of
   Warner 1995). The Doppler map is scaled so that, in velocity co-ordinates, 
   the M dwarf and white dwarf are at their correct respective positions 
   (and by implication the Roche lobes are as well). The vertical line from 
   the L1 point is the mass transfer stream in velocity co-ordinates. See 
   text for discussion of the significance of all these features.
   }
   \end{figure}  

A Doppler tomogram (Marsh \& Horne 1988) of the H$\alpha$ emission line
in the SAAO data was calculated using the code of Spruit (1998). The
results are shown in Fig. 16 which includes both velocity and spatial 
co-ordinates (the former on the lower horizontal and left vertical axes, 
the latter on the upper horizontal and right vertical axes). The origin
of the spatial co-ordinates is the white dwarf and that of the velocity
co-ordinates the binary centre of mass (depicted by a cross). The scales
were adjusted, using the binary parameters adopted from later analysis, 
so that unity in the spatial system corresponds to ${\rm (K_1 + K_2)\
sin\ i}$ in the velocity system. The centre of mass of the M dwarf and the 
white dwarf are thus at their correct position in both co-ordinate systems, 
and so are the Roche lobes. We are aware that this implies that the velocity 
of any specific feature in the Doppler map arises {\em only} from the binary 
rotation and no other velocity component is included (e.g. the velocity of
any streaming motions). Indeed, in velocity 
co-ordinates the mass transfer stream, as it moves away from the L1
point, quickly acquires ballistic velocities well in excess of the 
local binary rotation velocity and this is illustrated in Fig. 16
by the mass transfer stream whose velocity locus is the almost vertical
line from the L1 point. Checks on the velocity of the stream shows that
it never intersects the Roche lobe of the primary star.

With these considerations in mind, it is apparent that the majority of
the emission arises from the vicinity of both Roche lobes. Further, the
strongest emission component is centred on the Roche lobe of the M dwarf
and has no bias towards or away from its inner hemisphere facing the
white dwarf. This emission component corresponds to the line labelled 
`2' in Fig. 15.

Within the white dwarf Roche lobe, the emission {\em is} biased towards
the side facing the M dwarf and there are two distinct features at
velocity co-ordinates (Vx, Vy) = (-25, -75) and (75, -75) km s$^{-1}$.
The latter corresponds to the curve labelled `3' in Fig. 15. The
former feature does not fit the emission in the phase interval 0.0 to
0.25 as well as the curve labelled `1' in Fig. 15. The interpretation 
of these two distinct features in the Doppler map is not straightforward.
Neither occurs at a ``special" place in the binary: e.g. neither feature 
is associated with the white dwarf. Furthermore, it is clear that 
while there is no accretion disc in the system of the kind seen in
cataclysmic variables, there may be a {\em very} weak mass transfer
stream flowing from the L1 point. In this case, one might expect features 
in the Doppler map corresponding to where this stream collides with itself.
This could lead to an inviscid (i.e. dissipationless) disc which would be very
hard to detect (see Warner 1995 and references therein for a discussion of 
inviscid discs within the context of disc formation). If there is, there
might be a feature in the Doppler map corresponding to where the stream
collides with such a disc. The radius of an inviscid disc is shown 
in Fig. 16 by the small circle around the white dwarf (eq. 2.19 of Warner 
1995). One of the two distinct velocity features mentioned occurs near to, 
but not at, the intersection of the mass transfer stream with itself or 
the radius of an inviscid disc. The other feature is near to, but not
at the place where the ballistic stream slows down to low velocities at 
the edge of its Roche lobe before plunging back for a second passage 
around the white dwarf.

We caution against expecting too much from the Doppler tomogram: the 
technique does not allow for intrinsic variations in emission line strength 
due to optical thickness in the lines (which are clearly present); moreover, 
there is some variation in the maps produced by spectra from each half of the 
orbital cycle (not shown). We are impressed that the emission line behaviour
is not transient (it was seen in both the SAAO data on different occasions
and in the MSSSO data). We believe that these results indicate the presence 
of streaming motions in the white dwarf Roche lobe implying the existence of 
mass transfer, albeit at a very low level. There is undoubtedly accretion on
to the white dwarf from mass ejected by the M dwarf due to flares or its
wind, i.e. not via the L1 point. Such accretion would not cause specific 
and stable features in the Doppler map. Clearly more data with at least 
double the resolution in wavelength and binary phase are required to 
elucidate these issues further.

In the right hand panel of Fig. 15, it is worth drawing attention to the 
additional emission occurring at orbital phase 0.35 and positioned to the 
red of the most redward motion of the M dwarf. Although these are phase 
folded spectra, the MSSSO data have only 3-4 spectra per bin on average 
and examination of the data that contribute to this phase bin show a clear
triple structure to the emission line. Subsequent spectra showed the
additional peak following the motion of the M dwarf indicating that the
origin of the flare is close to that of the M dwarf and therefore
attributable to the flaring behaviour noted above. There is also evidence 
for similar behaviour (but with the additional velocity component occurring 
in the extreme blue wings of the line) in some of the SAAO spectra from 1993
Apr 19/20. Flaring in the H$\alpha$ emission line has also been seen by
Kawka et al. (2002).

The behaviour of the blue emission lines H$\beta$, H$\gamma$ and H$\delta$ 
was examined using the SAAO data from 1993 Feb and 1992 Apr 02/03, as well 
as the MSSSO data. The signal-to-noise in these data were generally much
worse than in the H$\alpha$ data shown in Fig. 15; despite this, the behaviour 
of these features was consistent with that seen in H$\alpha$: a two-component 
profile with obvious doubling at phase 0.25 but with one component much weaker 
half an orbital cycle later. 

The CaII K line's behaviour was somewhat different in that it never
showed doubling, was of lower velocity amplitude than the Balmer lines,
and showed a strong variation around the orbital cycle with peak strength
at orbital phase 0.5 and essentially disappearing at eclipse. This 
behaviour is strongly suggestive that this feature is formed on the
inner hemisphere of the M dwarf facing the white dwarf.

Finally, the $H\alpha$ emission line was used to provide a more accurate
estimate of the rotational velocity of the M dwarf than the absorption
line value derived above. This is justified as it is clear from the
Doppler tomography above that the velocity of the stronger component of 
the emission line is identical to that of the M dwarf. Fits of Doppler 
profiles to the emission line, and allowing for the instrumental resolution, 
yielded a value of V${\rm_{rot,2}\ sin\ i = 140 \pm 10}$ km s$^{-1}$.

   \begin{figure} 
   \begin{tabular}{c}
   \psfig{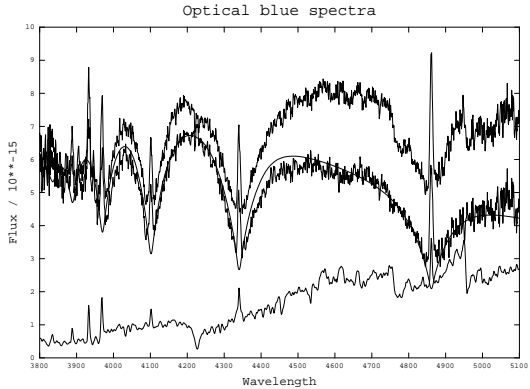} 
   \end{tabular}
   \caption[example]  
   { \label{fig:fig17}	   
   The top curve is the sum of the blue spectra from 1992 Apr 02/03 and
   1993 Feb. The bottom curve is the spectrum of the only M4.5eV star in
   the spectral atlas of Jacoby, Hunter \& Christian 1984. The middle
   smooth curve is the white dwarf model fit from the HST data; the
   remaining middle curve is the difference between the top and bottom
   curves.
   }
   \end{figure}  

\subsubsection{The blue spectra}

The blue spectra from 1992 Apr 02/03 and from 1993 Feb were selected
and summed together into 10 bins. Although the motion of the Balmer
emission lines was obvious, no sign of the orbital motion of the white
dwarf could be detected because the lines were too broad and the
signal-to-noise ratio insufficient. Instead, the data were simply summed 
together and fluxed using the lower resolution spectrum in Fig. 2. The 
result is shown in Fig. 17 (top curve). It is clear that this spectrum is 
contaminated not only by Balmer emission but also by absorption features 
from the M dwarf, especially at the red end. We had no observations of M 
dwarfs in this spectral region, we could find no publicly available models 
with features to match the observations, so we made use of the observed
spectrum of the M4.5eV star in the spectral atlas of Jacoby, Hunter
\& Christian (1984). Although its spectral type is 1 subclass later
than that of the M dwarf in EC13471--1258, it was satisfactory for
our purposes. A scaled version of this spectrum appears as the bottom
curve in Fig. 17. The middle smooth curve shows the model for the white
dwarf (T$_{\rm eff} = 14420$ K, log g = 8.34) along with the difference
between the top and bottom curves. Allowing for subtraction uncertainties
and instrumental effects, these two curves are a satisfactory match.

We tried to extract an independent estimate of effective temperature and
gravity from the subtracted spectrum. This procedure was difficult as the
emission cores had to be excluded and a fit to the continuum forced in
several places. We decided that no reliable, independent estimate of the
atmospheric parameters could be derived from this procedure. However, the
solution was consistent within uncertainties with the HST value.

\section{Analysis of the Binary}

The numerical results from the previous analyses are listed in Table
12 along with their errors. Error estimates for some quantities, e.g.
the eclipse analysis, are not shown as they are so small compared to
the other error estimates that they can safely be neglected. A subset
of these parameters allow derivation of the fundamental properties of 
the binary using the standard equations of spectroscopic binaries to 
derive the masses and radii of the individual components: 
\begin{eqnarray*}
{\rm a}              & = & {\rm P_{orb} (K_1 + K_2) / 2\pi}\\ 
{\rm M}              & = & {\rm 4\pi^2\ a^3 / G\ P_{orb}^2}\\ 
{\rm q}              & = & {\rm K_1 / K_2}\\
{\rm M_1}            & = & {\rm M / (1+q)}\\
{\rm M_2}            & = & {\rm q\ M_1}\\
{\rm f(R_1, R_2, i)} & = & 0\\ 
{\rm R_{lobe,2} / a} & = & {\rm 0.49q^{2/3}/(0.6q^{2/3}+ln(1+q^{1/3}))}
\end{eqnarray*}
where a is the binary separation, M is the sum of the masses of the
components, q is the ratio of the secondary to primary mass and
${\rm R_{lobe,2}}$ is the volume-equivalent Roche lobe radius of the
secondary (Eggleton 1983).

\begin{table}
\caption{Summary of results}
\begin{center}
\begin{tabular}{cccc}
Quantity & Value & Error & Units\\
         &       & Estimate \\    
\\
    P$_{\rm orb}$        &   13025.5 &             & s \\
 f(r$_1$/a, r$_2$/a, i)  &   Table 3 \\
 T${\rm _{eff,1}}$       &    14220  & $\pm$350    & K\\
       log g             &     8.34  & $\pm$0.22\\
       log Z             &    -1.5   & $\pm$0.5\\
     K$_1$ sin i         &     137   & $\pm$10     & km s$^{-1}$\\
     $\gamma_1$          &      61   & $\pm$10     & km s$^{-1}$\\
 V$_{\rm rot,1}$ sin i   &     400   & $\pm$100    & km s$^{-1}$\\
    K$_2$ sin i          &     266   & $\pm$6      & km s$^{-1}$ \\
     $\gamma_2$          &     -2    & $\pm$5      & km s$^{-1}$ \\
 V$_{\rm rot,2}$ sin i   &     140   & $\pm$10     & km s$^{-1}$\\
 T${\rm _{eff,2}}$       &    3100   & $\pm$75     & K\\
\end{tabular}
\end{center}
\end{table}

   \begin{figure} 
   \begin{tabular}{c}
   \psfig{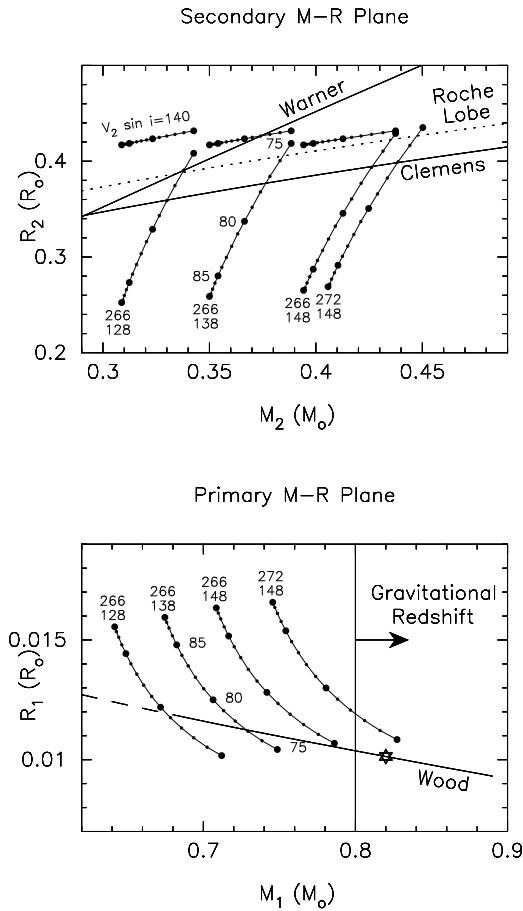} 
   \end{tabular}
   \caption[example]  
   { \label{fig:fig18}	   
   Mass-radius diagrams for the primary (lower panel) and secondary (upper)
   panel). The white dwarf mass-radius relation from Wood (1995) is shown
   in the lower panel labelled with `Wood'. Mass-radius relations for lower 
   main sequence stars are shown in the upper panel: equation 2.99 of Warner 
   (1995) is labelled `Warner' and that derived in Clemens et al. (1998), 
   based on Reid \& Gizis (1997), is labelled `Clemens'. Each 
   sloping line represents a solution for the binary using radial velocity 
   semi-amplitudes indicated by the pair of numbers at the end of the
   corresponding line. Each dot along each line represents a one degree
   change in binary inclination; one line in each panel is labelled with
   the numerical value of the inclination. The 6-sided star on the white
   dwarf mass-radius relation represents the estimate of log g from the
   model atmosphere fitting to the HST data. In the upper panel, the
   dotted line represents the limit where the mean density of the M dwarf
   would exceed the Roche lobe mean density associated with that orbital
   period, and would therefore overfill the lobe. The mass derived from 
   the gravitational redshift is 0.9 M$_\odot$; the vertical line in the 
   lower panel is the acceptable mass range derived from the 1-$\sigma$ 
   error estimate on the gravitational redshift. See the text for details 
   of other features and additional discussion.
   }
   \end{figure}  

Measurements in Table 12 show values for K$_1$~sin~i and K$_2$~sin~i,
not K$_1$ and K$_2$. As is well known in the analysis of spectroscopic
binaries, although a is only dependent on K$_1$~sin~i and K$_2$~sin~i, 
M is dependent on the cube of these quantities. In the following analysis,
therefore, uncertainties in the component masses and radii due to ignorance 
of the inclination and measurement errors in the radial velocity curves 
will be made explicit. The approach taken, therefore, was to try all integer
values of i between 75$^{\rm o}$ and 90$^{\rm o}$ for the optimal values 
of K$_1$~sin~i and K$_2$~sin~i in Table 12, and to repeat the calculations 
for perturbations on the velocity semi-amplitudes using the error estimates 
in Table 12. Mass-radius diagrams showing the results of these calculations 
for each component appear in Fig. 18. Several lines appear in these diagrams 
which will require some explanation before discussing their significance:

\begin{itemize}
\item Consider first the lower panel (the white dwarf mass-radius
      diagram) and within this panel the line labelled with the numbers
      266, 138. This line defines the locus of white dwarf mass-radius
      solutions for the optimal values of K$_1$~sin~i and K$_2$~sin~i 
      from Table 12. Different inclination angles are shown as filled
      circles on this line. Larger filled circles are shown every five
      degrees along with a label showing the numerical value of the
      inclination. This line extends from 90$^{\rm o}$ (at the top) to
      75$^{\rm o}$. 

\item Lines parallel to the one just discussed are shown and represent
      solutions for the values of K$_1$~sin~i and K$_2$~sin~i indicated
      at the top of the line. In the upper panel corresponding lines 
      are shown in the mass-radius diagram for the M dwarf. These loci
      will be referred to as ``dynamic solutions".

\item In the lower panel the line labelled `Wood' indicates the mass-radius
      relation for white dwarfs from the finite temperature, evolutionary
      models of Wood (1995). On this line is a 6-sided star which indicates
      the mass and radius of the white dwarf deduced from the value of log~g
      combined with the Wood mass-radius relation. The uncertainty on the
      value of log g is indicated by showing the mass-radius relation as
      a solid line. Beyond the 1-$\sigma$ error interval, the mass-radius 
      relation is shown as a dashed line.

\item We interpret the difference between the $\gamma$ velocities of the
      white dwarf and M dwarf as arising from the gravitational redshift
      of the white dwarf. This directly translates into a white dwarf mass
      of 0.9 M$_\odot$. The uncertainty on the difference allows a lower
      mass estimate and the allowable interval (at the 1-$\sigma$ level)
      is indicated by the vertical line in the lower panel with the left
      pointing arrow attached and labelled `Gravitational Redshift'.

\item As noted above, in the upper panel the dynamic solutions encompass
      a range of inclinations from 90$^{\rm o}$ to $\sim75^{\rm o}$. At 
      lower inclinations, the secondary would exceed its Roche lobe and
      are presumably excluded solutions. The parts of the dynamic solutions
      so excluded lie above the dotted line in the upper panel. This
      explains why the range of inclinations extended from 90$^{\rm o}$
      to only 75$^{\rm o}$. Note that the dotted line is essentially a
      geometric constraint, being based on the period-mean density
      relation for Roche lobe filling secondaries (see Warner 1995).

\item Mass-radius relations for lower main sequence stars are shown in the 
      upper panel: equation 2.99 of Warner (1995) is labelled `Warner' and 
      that derived in Clemens et al. (1998), based on Reid \& Gizis (1997), 
      is labelled `Clemens'. 
     
\item The remaining three almost horizontal lines in the upper panel, the
      leftmost of which is labelled `V$_{\rm rot,2}$', delineate a locus
      of M dwarf radii derived from the measurement of V$_{\rm rot,2}$~sin~i. 
      The dependency on sin~i is shown in the same manner as for the radial 
      velocity lines which also define the corresponding value of M$_2$. 
      Three lines are shown for three combinations of K$_1$~sin~i and 
      K$_2$~sin~i.

\end{itemize}

Assuming that M dwarf radii in excess of the Roche lobe radius are excluded,
the dynamic solutions representing the radial velocity semi-amplitudes and 
their $1-\sigma$ uncertainties, as well as the eclipse analysis, restrict 
the masses of the components to 0.59--0.83 and 0.30--0.45 M$_\odot$ for the 
white dwarf and M dwarf respectively. More precise values are available 
only if additional constraints are used.

The additional, purely observational constraints available are the loci
of the M dwarf rotational velocity in the upper panel, and the gravitational
redshift in the lower panel. The uncertainty on the M dwarf rotational
velocity is 7 per cent or 1.5 carets on the ordinate axis of the upper
panel. This restricts the range of allowable inclinations to less than 
77$^{\rm o}$. The gravitational redshift constrains the allowable range
for the white dwarf mass to be larger than 0.8 M$_\odot$. The dynamic
solutions impose a corresponding lower limit on the M dwarf mass to be
larger than 0.43 M$_\odot$.

Inclusion of the Wood (1995) white dwarf mass-radius relation restricts
the allowable inclinations to be smaller than 80$^{\rm o}$.

Inclusion of the mass-radius relations for the M dwarf leads to interesting
conclusions: (i) the relation in Warner (1995), which is typical of 
mass-radius relations from a number of sources, along with the assumption that 
the M dwarf radius does not overfill its Roche lobe, restricts the M dwarf 
mass to be smaller than 0.34 M$_\odot$. This is incompatible with the 
gravitational redshift constraint; (ii) Clemens et al. (1998) used the 
near infrared observations of Reid \& Gizis (1997) to derive a mass-radius 
relation for M dwarfs, discovering an abrupt change in slope of the 
mass-radius relation as well as incompatibility with the mass-radius 
relations of the best theoretical models (see Reid \& Hawley 2000 and 
references therein for a full discussion). The upper panel of Fig. 18
shows that the Clemens et al. relation, which diverges from the Warner
and other ``standard" relations at a secondary mass of 0.3 M$_\odot$, are 
compatible with the dynamical solutions for the M dwarf in EC13471--1258.

In arriving at a solution to the binary parameters, we attempted to
achieve consistency with: (i) the M dwarf not overfilling its Roche lobe; 
(ii) the atmospheric parameter determination and Wood mass-radius relation
for the white dwarf; (iii) the Clemens et al. mass-radius relation for the
M dwarf (bearing in mind that there is intrinsic scatter in this relation);
(iv) the V$_{\rm rot, 2}$ sin i determination for the M dwarf; (v) the
gravitational redshift measurement; (vi) the dynamical solutions. In 
addition, we believe that the M dwarf is essentially filling its Roche
lobe: evidence for this comes from the pre-eclipse dips (Fig. 8) without
accompanying flare, and the specific features seen in the Doppler tomogram.
If so, the solution will be on the dotted line in the upper panel of Fig.
18. Even if this is not the case, it is clear that the M dwarf is very
close to filling its Roche lobe.

Accordingly, as the best ``compromise" between all these constraints and
their errors, we adopt an inclination of $75.5\pm2.0^{\rm o}$, a white dwarf 
mass and radius of $0.78\pm0.04$ M$_\odot$ and $0.011\pm0.01$ R$_\odot$, and 
an M dwarf mass and radius of $0.43\pm0.04$ M$_\odot$ and $0.42\pm0.02$ 
R$_\odot$. We regard the error estimates as conservative. It is an
endorsement of stellar evolution observation and theory that all the 
relevant constraints can be satisfied simultaneously.

\subsection{Comparison with the results of Kawka et al.}

Recently, Kawka et al. (2002) have reported observations and analysis 
of EC13471--1258 along with two other DA+dM systems. In general, their
observational results broadly agree with those presented here. In the
following parentheses, the results of Kawka et al. appear first, those
presented in this paper second: the values of white dwarf log g 
(8.26, 8.34), T$_{\rm eff}$ (14080 K, 14220 K), and V$_{\rm 1,rot}$ sin i 
(140 km s$^{-1}$, 140 km s$^{-1}$) are all consistent within the errors 
(though we are sceptical of their small error of 0.05 on log g, given the 
analysis of the error on this quantity shown above). There is, however, 
discord in the M dwarf radial velocity semi-amplitude ($241\pm8.1$, $266\pm6$). 
We are inclined to prefer our value as it was derived from the absorption 
features of the M dwarf and is also consistent with the motion of the
stronger component of the emission lines. 

The analysis of Kawka et al. is somewhat flawed in that they use the
V sin i value to derive the absolute radius of the M dwarf without
dependence on sin i. They then use the duration of the white dwarf 
eclipse along with the absolute radius of the M dwarf to derive the
inclination (i.e. neglecting that this relationship also includes a,
the orbital separation). Fortuitously, their inclination of 73.5$^{\rm o}$
is reasonable. 

However, this inclination and, especially, the low value for the M dwarf 
radial velocity semi-amplitude yields a large value of 0.58 M$_\odot$ for
the M dwarf mass. This, along with the M dwarf radius of 0.42 R$_\odot$
can be excluded for two reasons: (i) the radial velocity semi-amplitude
for the white dwarf implied is about 185 km s$^{-1}$ (not 165 km s$^{-1}$ as 
claimed by Kawka et al.). This is excluded by the results in Table 8 and Fig. 
12.  Indeed, Kawka et al. derived a white dwarf radial velocity semi-amplitude 
of 155 km s$^{-1}$ but with large error (35 km s$^{-1}$), presumably because it 
seems from their fig. 11 that they used only one G140L HST spectrum from each 
orbit; (ii) the M dwarf mass and radius fall far from the locus of these
quantities, either theoretical or empirical, derived from a large sample
of M dwarfs (fig. 3 of Clemens et al. 1998; Reid \& Gizis 1997). In addition, 
the spectral type implied by the M dwarf mass is M0 or earlier (see Reid \& 
Hawley 2000) and discordant with the spectral type of M3.5-4 derived in 
this paper, or M2-M4 listed by Kawka et al. themselves. We believe that
the M dwarf mass of Kawka et al. should be disregarded.

\section{Discussion}

\subsection{Evolutionary status of the binary}

The canonical theory for the formation of cataclysmic variables involves
common envelope evolution enforcing dramatic orbital shrinkage with
a hot white dwarf and an M dwarf emerging from the envelope of the
AGB predecessor of the white dwarf. Following this, further orbital
shrinkage takes place via angular momentum loss by the M dwarf through
a weak wind with eventual contact. Further evolution of the semi-detached
cataclysmic variable requires ongoing angular momentum loss to ensure
ongoing Roche lobe contact by the M dwarf. These scenarios are discussed
in Warner (1995) and references therein.

There is also an accumulation of evidence that the mass transfer rate in
cataclysmic variables varies widely and perhaps cyclically on short time
scales. Mass transfer may even shut off and the system ``hibernates", a
term originated to explain the very existence of classical novae: in order 
to get a thermonuclear detonation with the vigour of those observed, the
mass accretion rate was required to to be very low immediately before the
nova explosion (Shara et al. 1986). The low mass accretion rate is now
no longer believed to be so important (Prialnik \& Kovetz 1995). More 
recent and direct evidence for variations in mass transfer rate is provided
by the very wide range of mass transfer rates at a given orbital period evident 
in diagrams such as fig. 9.8 of Warner (1995). One hypothesis to explain 
such mass transfer rate variations is irradiation of the M dwarf by the 
white dwarf. 

In terms of the above, therefore, EC13471--1258 is either a post common
envelope binary just at the phase where its M dwarf comes into contact 
with its Roche lobe for the first time, or it is a system that has been
in contact and is currently hibernating. We believe that two pieces of
evidence favour the latter interpretation: 

\begin{itemize}
\item the rotational velocity of the white dwarf: with V$_{\rm rot,1} =
      400 \pm 100$ km s$^{-1}$, the star rotates in 120 s. Such a rapid rotation
      rate implies that there has been an earlier phase of spin-up by mass
      transfer. Indeed, inspection of table 4 of Szkody et al. (2002a) shows
      that this rotation velocity is typical of cataclysmic variables. 
\item the behaviour of the H$\alpha$ emission line profile: although the
      interpretation of the Doppler tomogram is not clear, what is clear
      is that there are two or more components in the emission line and
      intrinsic variability as well, even in the component associated with
      the M dwarf. As the Balmer lines are optically thick, it is clear
      that the intrinsic variability can be understood by absorption by
      gas in the Roche lobe of the primary. The fact that the variation
      is a function of binary phase indicates that the absorption sites
      are fixed in the rotating reference frame of the binary and hence
      that there is a very weak mass transfer stream.
\end{itemize}
We therefore conclude that EC13471--1258 is a hibernating cataclysmic
variable.

Interestingly, the white dwarf temperature in EC13471--1258 is cool 
compared to white dwarfs in cataclysmic variables of similar orbital 
period. Indeed the temperature of this white dwarf is more typical of
systems below the period gap (fig. 3 of Szkody et al. 2002a). It is
close to the temperature of the pulsating white dwarf in GW Lib (Szkody
et al. 2002b), but shows no oscillations.

The white dwarf radius and model atmosphere imply a distance to the system 
of $48\pm5$ pc, well within the reach of ground-based parallax programs.
An accurate parallax would enable a more stringent test of the white 
dwarf mass-radius relationship than presented in this paper. An absolute
V magnitude of the white dwarf was derived: 11.74.

\subsection{The M dwarf}

The analysis of the M dwarf presented in this paper shows that its properties 
are compatible with field stars of similar spectral type. This M dwarf is of 
particular relevance to studies of M dwarfs of the kind conducted by Reid 
\& Gizis (1997) and Clemens et al. (1998) as its properties (V-I colour, 
mass and radius) place it in the segment of the M dwarf sequence where an 
abrupt change in slope of many stellar parameters occurs. This abrupt change 
in slope is discussed in Clemens et al. (1998) and Reid \& Gizis (1997) to 
which the reader should turn for further information.

The M dwarf exhibits the properties expected of a secondary star in
a cataclysmic variable: it is extremely active, presumably because of its
very short rotation period. The erratic behaviour of the binary period is
also typical of cataclysmic variables and presumably arises from magnetic
cycling in the M dwarf (e.g. Warner 1995 and references therein). If flares 
of the kind described in this paper occurred on the secondaries of cataclysmic 
variables with higher mass transfer rate, they would be indistinguishable 
from the flickering in the system.

Compatibility with the Clemens et al. mass-radius relation implies that
the M dwarf is 250 K hotter than the 3100 K derived above: see fig. 4.13
of Reid \& Hawley (2000). As they discuss, the temperature scale for M
dwarfs is uncertain by this amount. Resolution of this discrepancy must
await further developments in M dwarf model atmosphere research.

A puzzling feature of the photometry of the M dwarf is the fact that the
ellipsoidal variation is not constant and is often asymmetrical with 
respect to the conjunction of the two stars. When this happens, its 
minimum occurs slightly before orbital phase 0. If this behaviour could
be attributed to star spots on the M dwarf, it would not be puzzling.
However, spots on other chromospherically active stars migrate. In this
case, 10 years of eclipse monitoring has shown that the asymmetry is
either absent or always in the same sense. 

Using the distance derived from the last subsection, the absolute V magnitude 
of the M dwarf is 11.82.

\subsection{Chemical composition of the binary components}

The chemical compositions of the binary components are different: that 
of the M dwarf is normal solar, that of the white dwarf is 1/30 solar.
As has been well known for decades, the chemical composition of the 
photospheres of typical single DA white dwarfs is extremely metal deficient 
($<10^{-6}$) due to gravitational settling of heavy elements in the very high 
surface gravity of the white dwarf. The presence of metals in the white dwarf 
in EC13471--1258, therefore, is due to mass accretion from the nearby M dwarf 
(whether this mass transfer is from the wind of the M dwarf or Roche lobe 
overflow is not relevant to this discussion). Metals are also seen in the
white dwarf primaries of cataclysmic variables where sub-solar metal 
abundances are common (see table 4 of Szkody et al. 2002a). The abundances
nevertheless vastly exceed the abundances in single white dwarfs.

If the chemical composition difference is in steady state, and the mass 
accretion rate on to the white dwarf can be estimated, the diffusion 
coefficients for the chemical species visible in the HST spectra can be 
deduced. This is beyond the scope of this paper but is clearly an 
interesting avenue for future work as this system provides a natural 
diffusion laboratory.

\subsection*{Acknowledgements}

DOD thanks Ed Sion for a heroic attempt to observe this star with IUE. We 
are grateful to Xavier Delfosse for supplying the radial velocity of GJ381, 
to Hans Ritter, Andrew King, Boris Gansicke, Brian Warner and Mike Shara
for helpful conversations, to Henk Spruit for the use of his Doppler 
tomography software, and to Steve Potter for help with its implementation. 
This paper has had an even longer gestation period than that of O'Brien, 
Bond \& Sion (2001); DOD thanks Kawka and collaborators for what some of 
his co-authors and colleagues tried for so long to exert: pressure to 
finish the paper.

\bsp

\end{document}